\documentclass[aps,prb,twocolumn,showpacs,nofootinbib,longbibliography]{revtex4-1}
\usepackage{amsmath,latexsym}
\usepackage{amssymb}

\usepackage{graphicx}
\usepackage{dcolumn}
\usepackage{bm}
\usepackage{xcolor}
\usepackage[%
  colorlinks=true,
  urlcolor=blue,
  linkcolor=blue,
  citecolor=blue
]{hyperref}
\usepackage{mathptmx}
\usepackage{tabularx}
\usepackage{etoolbox}

\newcommand{\nn}{\nonumber}
\newcommand{\beq}{\begin{eqnarray}}
\newcommand{\eeq}{\end{eqnarray}}

\makeatletter
\let\cat@comma@active\@empty
\makeatother
\begin{document}

\title{
Topology and symmetry of surface Majorana arcs in cyclic superconductors
}

\author{Takeshi Mizushima}
\email{mizushima@mp.es.osaka-u.ac.jp}
\affiliation{Department of Materials Engineering Science, Osaka University, Toyonaka, Osaka 560-8531, Japan}
\author{Muneto Nitta}
\email{nitta@phys-h.keio.ac.jp}
\affiliation{Department of Physics, and Research and Education Center for Natural Sciences,
Keio University, Hiyoshi 4-1-1, Yokohama, Kanagawa 223-8521, Japan}

\date{\today}

\begin{abstract}
We study the topology and symmetry of surface Majorana arcs in superconductors with nonunitary ``cyclic'' pairing. Cyclic $p$-wave pairing may be realized in a cubic or tetrahedral crystal, while it is a candidate for the interior $^3P_2$ superfluids of neutron stars. The cyclic state is an admixture of full gap and nodal gap with eight Weyl points and the low-energy physics is governed by itinerant Majorana fermions. We here show the evolution of surface states from Majorana cone to Majorana arcs under rotation of surface orientation. The Majorana cone is protected solely by an accidental spin rotation symmetry and fragile against spin-orbit coupling, while the arcs are attributed to two topological invariants: the first Chern number and one-dimensional winding number. Lastly, we discuss how topologically protected surface states inherent to the nonunitary cyclic pairing can be captured from surface probes in candidate compounds, such as U$_{1-x}$Th$_{x}$Be$_{13}$. We examine tunneling conductance spectra for two competitive scenarios in U$_{1-x}$Th$_{x}$Be$_{13}$---{\it the degenerate $E_u$ scenario and the accidental scenario.}
\end{abstract}


\maketitle

\section{Introduction}

The intense studies on anisotropic superfluidity and superconductivity in condensed matter and nuclear matter were initiated by the discovery of spin-triplet ($S=1$), $p$-wave ($L=1$) superfluidity in $^3$He and the prediction of $^3P_2$ superfluidity in the dense core of neutron stras, respectively.~\cite{osheroffPRL72,tamagaki,hoffberg} The liquid $^3$He, which behaves as isotropic Fermi liquid, preserves the separate rotation symmetry in spin and orbital spaces, $G={\rm SO}(3)_S\times {\rm SO}(3)_L$. In $^3$He-B, which occupies the almost region of the superfluid phase diagram, the pairing maintains the total angular momentum $J = S+ L= 0$ and spontaneously breaks the spin-orbit symmetry.~\cite{vollhardt,leggettRMP75} In contrast, in dense neutron matter, a short-range attractive spin-triplet $p$-wave interaction originates in a strong spin-orbit force mediated by the exchange of vector mesons in nuclei and the existence of a repulsive core in the $^1S_0$ channel prevents the formation of conventional $s$-wave pairing.~\cite{tamagaki,hoffberg,wolfAJ66,takatsukaPTP72,fujitaPTP72,richardsonPRD72} The Cooper pairs glued by the strong spin-orbit force preserve the total angular momentum $J=2$, which are referred as to $^3P_2$ states.

\begin{figure}[t]
\includegraphics[width=80mm]{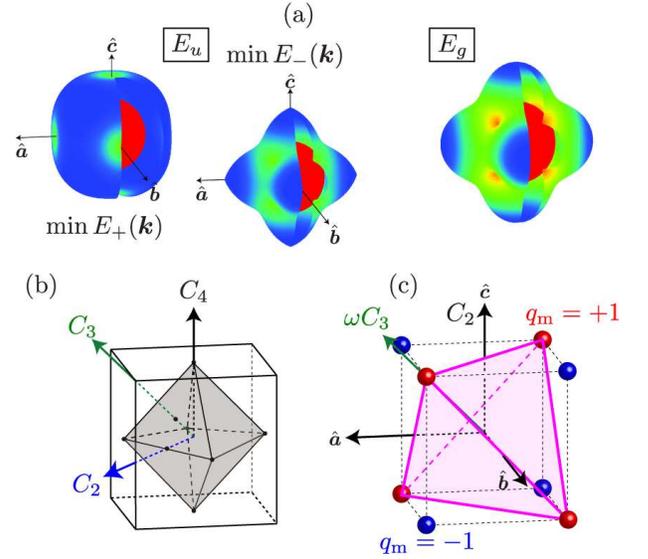}
\caption{(Color online) (a) Gap structure of the cyclic $p$-wave ($E_u$) and $d$-wave ($E_g$) states, where the former is composed of the fully gapped ($E_+({\bm k})$) and gapless ($E_-({\bm k})$) bands. (b) Discrete rotation symmetries in the $O_h$ symmetry group, which contains six $C_4$ axes and six $C_2$ axes in the horizontal plane, and eight $C_3$ axes. (c) Configuration of eight Weyl points in the cyclic $p$- and $d$-wave states, where each node is characterized by the monopole charge, $q_{\rm m}=\pm 1$. The gap function has three $C_2$ axes and four $\omega C_3$ axes in addition to the $\mathcal{C}$ and $\mathcal{P}$ symmetries, where $\omega C_3$ stands for the $C_3$ symmetry compensated by the $\omega =e^{i2\pi/3}$ phase rotation.
}
\label{fig:cyclic}
\end{figure}

$^3P_2$ superfluid phases include uniaxial/biaxial nematic phases, the ferromagnetic phase, and the cyclic phase.~\cite{richardsonPRD72,merminPRA74,saulsPRD78,saulsphd,masudaPRC16} The $^3P_2$ order parameter is represented by the the second-rank, traceless, and symmetric tensor, $A_{\mu i}$, which transforms as a vector with respect to index $\mu = x,y,z$ and under spin rotations, and, separately, as a vector with respect to index $i = x,y,z$ under orbital rotations. 
The nematic phases are represented by $A_{\mu i}=\Delta [\hat{a}_{\mu}\hat{a}_i + r \hat{b}_{\mu}\hat{b}_i -(1+r) \hat{c}_{\mu}\hat{c}_i]$ where $r\in [-1,-1/2]$ and $(\hat{\bm a}, \hat{\bm b}, \hat{\bm c})$ is an orthonormal triad. The uniaxial nematic state at $r=-1/2$ is fully gapped, while the biaxial state in $r\neq -1/2$ has nodal points. All the states are categorized into DIII topological class and their low energy physics is governed by two-dimensional helical Majorana fermions residing on the surface.~\cite{mizushimaPRB17} In contrast, the cyclic phase is the nonunitary state with the order parameter
\beq
A^{\rm cyclic}_{\mu i} = \Delta \left[
\hat{a}_{\mu}\hat{a}_i + \omega\hat{b}_{\mu}\hat{b}_i + \omega^2\hat{c}_{\mu}\hat{c}_i
\right],
\label{eq:Acyclic}
\eeq
where $\omega^3=1$. As shown in Fig.~\ref{fig:cyclic}(a), the quasiparticle gap structure is an admixture of the ful gap and nodal gap. Bogoliubov quasiparticles around nodal points behave as three-dimensional Majorana fermions and the nontrivial Berry curvature brings about characteristic surface states.~\cite{venderbosPRB16,kozii16,mizushimaPRB17} 



In addition to neutron stars, nematic and cyclic states can also be realized in odd-parity superconductors as the two-dimensional irreducible representation ($E_u$) of the cubic ($O_h$) point group symmetry~\cite{volovikJETP85,uedaPRB85,ozaki,sigrist} and tetrahedral ($T_h$) point group symmetry.~\cite{sergienkoPRB04} 
Indeed the possibility of nonunitary superconductivity has recently been argued in heavy fermion compounds, such as the filled skutterudite superconductor PrOs$_4$Sb$_{12}$~\cite{ichiokaJPSJ03,asanoPRB03,alrubPRB07,kozii16} and uranium compound U$_{1-x}$Th$_x$Be$_{13}$.~\cite{shimizu17} Understanding their gap and topological structure may be fed back to the interior $^3P_2$ superfluids of neutron stars.

The superconducting gap symmetry and the multiple superconducting phase transition in U$_{1-x}$Th$_x$Be$_{13}$ have been longstanding unsolved issues.~\cite{sigrist} The superconducting transition temperature, which is $T_{\rm c}\sim 0.85$K at $x=0$, shows non-monotonic behavior as dopant $x$ increases.~\cite{ottPRB85,fulde,scheidtPRB98,kromerPRL98,kromerPRB00} For $x <  0.019$, $T_{\rm c}$ decreases linearly with increasing $x$, while it shows a dome-like maximum of $T_{\rm c}$ in a narrow range of $0.019< x < 0.045$, which is referred to as $T_{\rm c1}(x)$. The local maximum of $T_{\rm c1}(x)$ appears at $x\sim 0.03$. In $0.019< x < 0.045$, another phase transition occurs at $T_{\rm c2}(x)$ ($< T_{\rm c1}(x)$). According to zero-field $\mu$SR experiment,~\cite{heffnerPRL90} the phase in $T<T_{\rm c2}$ breaks the time reversal symmetry, while the phase in $x< 0.019$ does not. Recent heat capacity and magnetization measurements at $x=0.03$ further indicate that $T_{\rm c2}$ is the second transition to a different superconducting state.~\cite{shimizu17} One of possible scenarios to resolve the issues is the {\it accidental} scenario, where multiple order parameters are assumed to belong to different irreducible representation of the $O_h$ group.~\cite{sigristPRB89,sigrist} Another senario is the odd-parity $E_u$ state.~\cite{shimizu17} This suggests the cyclic state in $T<T_{\rm c2}$, biaxial nematic states in $T_{\rm c2}<T<T_{\rm c1}$ for $0.019< x < 0.045$, and the uniaxial nematic state for $x< 0.019$.

In this paper, we clarify the topological aspect of nonunitary cyclic superconductors. The cyclic $p$-wave state hosts both three-dimensional Majorana fermions~\cite{venderbosPRB16,kozii16,mizushimaPRB17} emerging from the bulk Weyl points and surface Majorana fermions as a reflection of nontrivial topology in the bulk. We show that changing surface orientation leads to the evolution of surface bound states from gapless Majorana cone to Majorana arcs. The former is protected solely by accidental spin rotation symmetry and may be sensitive to perturbation with the broken symmetry, such as the Rashba spin-orbit coupling on surface. In contrast, Majorana arcs originate in two different types of topological invariants: the first Chern number and one-dimensional winding number, where the latter is attributed to the combination of the time-reversal symmetry and mirror reflection symmetry.
The evolution of surface Fermi arcs in cyclic $d$-wave states ($E_g$) has been argued in the context of Andreev bound states.~\cite{ishikawaJPSJ13} It turns out that the topology and symmetry of Majorana arcs in $E_u$ is essentially different from those of the $E_g$ state. We demonstrate that the evolution of surface Majorana fermions from a cone shape to arcs gives rise to the evolution of tunneling conductance from a split peak structure to zero bias conductance peak. Understanding topologically protected surface states inherent to the nonunitary cyclic pairing may provide a possible way to determine the gap symmetry of U$_{1-x}$Th$_x$Be$_{13}$ through surface probes.

%


This paper is arranged as follows. In Sec.~II, we clarify the connection between the gap structure and Berry curvature of the cyclic phase in the momentum space. The low energy Bogoliubov quasiparticles are comprised of single-species Weyl fermions with tetrahedral symmetry. 
In Sec.~III, based on numerical results on the angle-resolved surface density of states, we clarify the symmetry and topology of zero energy surface states in cyclic superconductors. We introduce two different types of one-dimensional winding numbers associated with  order-two discrete symmetries. The evolution of surface Majorana arcs with respect to surface orientation angles are discussed on the basis of the Chern number and winding numbers. Furthermore, in Sec.~\ref{sec:tunnel}, we present the surface density of states and tunneling conductance in U$_{1-x}$Th$_x$Be$_{13}$ superconducting junctions for various surface orientations and argue their connection with the evolution of surface Majorana fermions. The final section is devoted to conclusion and discussion. The framework of the quasiclassical theory is summarized in Appendix.

\section{Cyclic states in cubic symmetry}
\label{sec:cyclic}


The low energy physics of bulk superconductors are determined by the second quantized Hamiltonian, 
\beq
\mathcal{H} = E_0 + \frac{1}{2}\sum _{\bm k}{\bm c}^{\dag}({\bm k})\mathcal{H}({\bm k}){\bm c}({\bm k}), 
\label{eq:H0}
\eeq
where $E_0$ is the constant and ${\bm c}^{\dag}({\bm k})\!=\![c^{\dag}_{\uparrow}({\bm k}),c^{\dag}_{\downarrow}({\bm k}),c_{\uparrow}(-{\bm k}),c_{\downarrow}(-{\bm k})]$ is the creation and annihilation operators of electrons in the Nambu space. The Bogoliubov-de Gennes (BdG) Hamiltonian density is given by
\beq
\mathcal{H}({\bm k}) = \left( 
\begin{array}{cc}
\varepsilon({\bm k}) & \Delta({\bm k}) \\
-\Delta^{\ast}(-{\bm k}) & - \varepsilon^{\rm T}(-{\bm k}) 
\end{array}
\right).
\label{eq:H}
\eeq
We here suppose that the $2\times 2$ single-particle Hamiltonian density, $\varepsilon({\bm k})$, preserves $O_h$ crystalline symmetry when external fields are absent. The $2\times 2$ superconducting order parameter, $\Delta({\bm k})$, is decomposed into spin singlet scalar component $\psi({\bm k})$ and triplet vectorial components ${\bm d}({\bm k})$ as 
\beq
\Delta ({\bm k}) = i\sigma _b \psi({\bm k}) + i{\bm \sigma}\cdot{\bm d}({\bm k})\sigma _b.
\label{eq:delta}
\eeq 
The quasiparticle excitation energy at zero fields is given by diagonalizing Eq.~\eqref{eq:H} as
\beq
E_{\pm}({\bm k}) = \sqrt{\varepsilon^2_0({\bm k}) + |{\bm d}({\bm k})|^2 \pm |{\bm d}({\bm k})\times {\bm d}^{\ast}({\bm k})|},
\label{eq:epm}
\eeq
for spin triplet pairing and $E_{\pm}({\bm k})\!=\!\sqrt{\varepsilon^2_0({\bm k}) + |{\psi}({\bm k})|^2}$ for spin singlet pairing,
where $\varepsilon _0 ({\bm k})\!=\! \frac{1}{2}{\rm tr}\varepsilon({\bm k})$. Wave numbers and spin Pauli matrices are denoted as ${\bm k}=k_a\hat{\bm a} + k_b\hat{\bm b} + k_c{\bm c}$ and ${\bm \sigma}=\sigma _1\hat{\bm a} + \sigma _2\hat{\bm b} + \sigma _3{\bm c}$, respectively, in the basis of crystal coordinates, $(\hat{\bm a},\hat{\bm b},\hat{\bm c})$. In this paper, we set $\hbar=k_{\rm B}=1$. ${\bm \sigma}$ (${\bm \tau}$) is the Pauli matrices in spin (Nambu) space ($\mu = 1,2,3$) and $\tau _0$ is the unit matrix in the Nambu space. The repeated Greek and Roman indices imply the sum over $x$, $y$, and $z$.

\subsection{Discrete symmetries}

Let us summarize the fundamental discrete symmetries of Eq.~\eqref{eq:H} which are relevant to topological invariants: the particle-hole (${\rm C}$), time-reversal (${\rm T}$), inversion ($P$), and $n$-fold rotation ($C_n$) symmetries. These symemtries guarantee that the Hamiltonian in Eq.~\eqref{eq:H0} is invariant under the transformation of fermions with the momentum (${\bm k}$) and spin ($a,b=\uparrow,\downarrow$), 
${\rm C} c_{a}({\bm k}) {\rm C}^{-1} = \Xi _{ab}c^{\dag}_{b}(-{\bm k})$ ($\Xi \equiv \tau _x$ for odd parity pairing and $\Xi \equiv i\tau _2$ for even parity pairing), 
${\rm T} c_{a}({\bm k}) {\rm T}^{-1} = \Theta _{ab}c^{\dag}_{b}(-{\bm k})$ ($\Theta \equiv i\sigma _2$), 
$P c_{a}({\bm k}) P^{-1} = c_{a}(-{\bm k})$, 
and $C_n c_{a}({\bm k}) C_n^{-1} = U_{ab}(\hat{\bm n},\varphi _n)c_{b}(-{\bm k})$, where $\varphi _n\equiv 2\pi /n$ denotes the $n$-fold rotation angle and the ${\rm SU}(2)$ rotation matrix $U(\hat{\bm n},\varphi _n)$ represents a $n$-fold rotation of spin $1/2$ about $\hat{\bm n}$ axis and $R_n$ is the corresponding ${\rm SO}(3)$ matrix. 

The particle-hole symmetry (PHS) requires the BdG Hamiltonian density $\mathcal{H}({\bm k})$ to hold the relation 
\beq
\mathcal{C}\mathcal{H}({\bm k})\mathcal{C}^{-1}=-\mathcal{H}(-{\bm k}), 
\label{eq:phs}
\eeq
with $\mathcal{C}=\Xi K$, where $K$ is the complex conjugation operator. In addition, the time reversal symmetry (TRS) and inversion symmetry lead to
\begin{gather}
\mathcal{T}\mathcal{H}({\bm k})\mathcal{T}^{-1}=\mathcal{H}(-{\bm k}),
\label{eq:trs} \\
\mathcal{P}\mathcal{H}({\bm k})\mathcal{P}^{-1}=\mathcal{H}(-{\bm k}),
\label{eq:inversion}
\end{gather} 
with the time-reversal operator $\mathcal{T}=\Theta K$. The TRS guarantees that ${\bm d}({\bm k})$ and $\psi({\bm k})$ are real. The ``inversion'' operator is given by $\mathcal{P}=\tau _3$ for odd parity pairing and $\mathcal{P}=\tau _0$ for even parity pairing. For the odd parity case, the $\mathcal{P}$ operator contains the $\pi$ phase rotation of $\Delta$ that compensates the sign change of $\Delta$ induced by the inversion ${\bm k}\mapsto-{\bm k}$. The $n$-fold rotation symmetry associated with the point group symmetry of crystals is given as
\beq
\mathcal{U}_n(\hat{\bm n})\mathcal{H}({\bm k})\mathcal{U}^{\dag}_n(\hat{\bm n})=\mathcal{H}(R_n{\bm k}),
\label{eq:rot}
\eeq
where $\mathcal{U}_n(\hat{\bm n}) \equiv U(\hat{\bm n},\varphi _n)\oplus U^{\ast}(\hat{\bm n},\varphi _n)$ is the ${\rm SU}(2)$ matrix extended to the Nambu space and $R_n\equiv R(\hat{\bm n},\varphi)$ is the $n$-fold rotation matrix about $\hat{\bm n}$.
This requires that the diagonal and off-diagonal block matrix obeys the relations, $U_n\varepsilon({\bm k})U^{\dag}_n = \varepsilon (R_n{\bm k})$ and $U_n\Delta ({\bm k})U^{\rm T}_n = \Delta (R_n{\bm k})$. 

As displayed in Fig.~\ref{fig:cyclic}(b), the $O_h$ symmetry group possesses six $C_4$ rotations about the $\hat{\bm a}$, $\hat{\bm b}$, $\hat{\bm c}$ axes, six $C_2$ rotations in the $\hat{\bm a}$-$\hat{\bm b}$ plane, and eight $C_3$ rotations. We notice that owing to the presence of the inversion symmetry, the $C_2$ rotations are accompanied by the mirror reflection symmetry, 
\beq
\mathcal{M} \mathcal{H}({\bm k})\mathcal{M}^{-1} = \mathcal{H}(-R_2{\bm k}).
\label{eq:mirror}
\eeq
where the mirror reflection planes are normal to the $C_2$ rotation axes. 
The mirror reflection operator in the Nambu space is constructed from a combination of the $C_2$ rotation and inversion symmetries as $\mathcal{M} \equiv \mathcal{U}_2(\hat{\bm n})\mathcal{P} = M\oplus (-M^{\ast})$ for odd parity pairing. The operator $M\equiv -i{\bm \sigma}\cdot\hat{\bm n}$ stands for the mirror reflection that flips the momentum and spin as ${\bm k}\mapsto -R_2{\bm k} = {\bm k}-2\hat{\bm n}(\hat{\bm n}\cdot{\bm k})$ and ${\bm \sigma}\mapsto - {\bm \sigma}+2\hat{\bm n}(\hat{\bm n}\cdot{\bm \sigma})$, where $\hat{\bm n}$ characterizes a normal vector in the mirror reflection plane. In Sec.~\ref{sec:arc}, we will demonstrate that the mirror reflection symmetry is indispensable for understanding the evolution of surface Majorana arcs in odd-parity cyclic states. 


\subsection{Cyclic $p$- and $d$-wave states}

In this paper, we mainly consider the broken time reversal state in cubic crystals. According to the group theoretic classification under cubic crystalline symmetry,~\cite{volovikJETP85,ozaki,uedaPRB85,sigrist} there are two-dimensional irreducible representations of the $O_h$ symmetry group: $E_g$ and $E_u$ representations for even parity and odd parity states, respectively. The $E_u$ irreducible representation possesses the following two basis functions,
\begin{align}
{\bm \Gamma}^{E_u}_1({\bm k}) &= \frac{1}{\sqrt{2}}(2\hat{\bm c} \hat{k}_c
-\hat{\bm b}\hat{k}_b - \hat{\bm b}\hat{k}_b), \label{eq:gamma1} \\
{\bm \Gamma}^{E_u}_2({\bm k}) &= \sqrt{\frac{3}{2}}(\hat{\bm a}\hat{k}_a-\hat{\bm b}\hat{k}_b).
\label{eq:gamma2}
\end{align}
The odd-parity component of the superconducting gap in Eq.~\eqref{eq:delta} is then expanded in terms of these basis as
\beq
{\bm d} ({\bm k}) = \eta ^{E_u}_1 {\bm \Gamma}^{E_u}_1({\bm k})
+ \eta ^{E_u}_2 {\bm \Gamma}^{E_u}_2({\bm k}),
\label{eq:dvec}
\eeq
with complex variables $(\eta^{\Gamma}_1,\eta^{\Gamma}_2)$. 
The cyclic state in the $E_u$ representation is obtained as the chiral pairing with broken time-reversal symmetry, 
$(\eta^{E_u}_1,\eta^{E_u}_2) = (1,i)$.
The ${\bm d}$-vector is then recast into
\beq
{\bm d}({\bm k}) = \Delta \left( 
\hat{\bm a}\hat{k}_a + \omega \hat{\bm b}\hat{k}_b + \omega^2 \hat{\bm c}\hat{k}_c
\right),
\label{eq:dcyclic}
\eeq
with $\omega^3\!=\! 1$. Introducing the tensor representation, $d_{\mu}(\hat{\bm k}) = A_{\mu i}\hat{k}_{i}$, one finds that Eq.~\eqref{eq:dcyclic} is equivalent to the cyclic order parameter~\eqref{eq:Acyclic} in $^3P_2$ superfluids. 

As shown in Fig.~\ref{fig:cyclic}(c), the cyclic state spontaneously beaks the $O_h$ symmetry into the tetrahedral symmetry which has three $C_2$ axes along $\hat{\bm a}$, $\hat{\bm b}$, and $\hat{\bm c}$ and four $C_3$ axes accompanied by the $\omega = e^{i2\pi/3}$ phase rotation. The tetrahedron has three mirror reflection planes that contains the $\hat{\bm a}$, $\hat{\bm b}$, and $\hat{\bm c}$ axes and other mirror reflection symmetries are spontaneously broken. 


For $\Gamma = E_g$, the basis functions $[\psi^{E_g}_{1}({\bm k}),\psi^{E_g}_2({\bm k})]$ are obtained from Eqs.~\eqref{eq:gamma1} and \eqref{eq:gamma2} by replacing $(\hat{\bm a},\hat{\bm b},\hat{\bm c})$ to $(\hat{k}_a,\hat{k}_b,\hat{k}_c)$. The cyclic $d$-wave state is defined as $\psi^{E_g} _1 ({\bm k}) + i\psi ^{E_g}_2 ({\bm k})$, which can be recast into~\cite{ishikawaJPSJ13}
\beq
\psi ({\bm k}) = \Delta \left( 
\hat{k}^2_a + \omega \hat{k}^2_b + \omega^2 \hat{k}^2_c
\right).
\eeq
Similarly to the cyclic $p$-wave state, the time-reversal symmetry broken $E_g$ state possesses eight point nodes and maintains the tetrahedral symmetry. The gap structure is displayed in Fig.~\ref{fig:cyclic}(a).


The Ginzburg-Landau free energy functional for the two dimensional representations can be written with the coefficients $\beta _1$ and $\beta _2$ as~\cite{sigrist}
\beq
\mathcal{F}[\eta _m,\eta^{\ast}_m] = \alpha|{\bm \eta}|^2
+ \beta _1|{\bm \eta}|^4
+ \beta _2(|{\bm \eta}\cdot{\bm \eta}|^2-|{\bm \eta}|^4),
\eeq
where $\alpha(T)\!\propto\!T_{\rm c}-T$ and ${\bm \eta}\!=\!(\eta_1,\eta_2)^{\rm T}$. The time reversal broken cyclic phase with $(\eta_1,\eta_2)\propto (1,i)$ can be realized for $\beta _2 /\beta _1 < 0$, while the region $\beta _2/\beta _1 > 0$ is favored by time reversal invariant unitary states with $(\eta_1,\eta_2) = (\cos\theta,\sin\theta)$. The unitary states correspond to highly degenerate minima of $\mathcal{F}$ with respect to $\theta$. In the context of the $^3P_2$ superfluids which are expected to be realized in the inner core of neutron stars, the ordered state at $\theta =0$ is refereed to as the uniaxial nematic phase, while the biaxial nematic phase at $\theta=\pi/2$ is invariant under the the dihedral-four $D_4$ symmetry. The intermediate $\theta$ holds the dihedral-two $D_2$ symmetry. 

All the time reversal invariant $^3P_2$ superfluids with $\mathcal{T}^2\!=\! -1$ and $\mathcal{C}^2\!=\! + 1$ are categorized to the class DIII in the topological table.~\cite{schnyderPRB08} The topological structure of the $D_4$ biaxial nematic state which has a PHS pair of point nodes is characterized by the $\mathbb{Z}_2$ topological number.~\cite{mizushimaPRB17} This is equivalent to the topological structure of the planar state,~\cite{satoPRB10} the $E_{1u}$ state in UPt$_3$,~\cite{tsutsumiJPSJ12-2,tsutsumiJPSJ13,mizushimaPRB14} and the $E_u$ state in Cu$_x$Bi$_2$Se$_3$.~\cite{fuPRL10,satoPRB10} Since the uniaxial and $D_2$ biaxial nematic states are fully gapped, their topological structures are equivalent to those of the superfluid $^3$He-B~\cite{schnyderPRB08,volovikJETP09v2,mizushimaJPCM15,mizushimaJPSJ16} and the $A_{1u}$ state in Cu$_x$Bi$_2$Se$_3$.~\cite{fuPRL10,satoPRB10,sasaki15}

\subsection{Weyl fermions and Berry curvature}
\label{sec:weyl}

The nonunitary state has two distinct energy branches. In Fig.~\ref{fig:cyclic}(a), we display the gap structures, $\min E_{\pm}({\bm k})$, where the upper branch $E_+({\bm k})$ is fully gapped, and the lower branch $E_-({\bm k})$ has eight Fermi points. The tetrahedral symmetry guarantees that the four of Fermi points reside on four vertices of the tetrahedron, 
\beq
{\bm k}_{\rm node} = \{{\bm k}_0,C_{2,a}{\bm k}_0, C_{2,b}{\bm k}_0,C_{2,c}{\bm k}_0\}, 
\eeq
where one point node exists at the $(111)$ direction, ${\bm k}_0\!=\! k_{\rm F}(1,1,1)/\sqrt{3}$ (see Fig.~\ref{fig:cyclic}(b)). In addition, Eq.~\eqref{eq:phs} implies that the point nodes, which obeys $\det \mathcal{H}({\bm k}_{\rm node})=0$, must appear as a PHS pair in the ${\bm k}$-space, $\mathcal{C}\mathcal{H}({\bm k}_{\rm node}) \mathcal{C}^{-1}=-\mathcal{H}(-{\bm k}_{\rm node})$.

Let $S$ be a small surface enclosing a Weyl point in the ${\bm k}$ space and ${\bm s}$ be a normal vector to $S$. We here define the Chern number or the monopole charge on $S$ by
\beq
q_{\rm m} = \frac{1}{2\pi}\int _S d{\bm s} \cdot{\bm \Omega}_-({\bm k}),
\label{eq:monopole}
\eeq
where the Berry curvature in the occupied states of the $n$-th band is obtained from the eigenvectors of the BdG Hamiltonian, $| u_n({\bm k})\rangle$,
$[{\bm \Omega}_n({\bm k})]_{\mu} = i \epsilon _{\mu \nu \eta} \langle \partial _{k_{\nu}}u_n({\bm k}) | 
\partial _{k_{\eta}}u_n({\bm k})\rangle $.
The monopole charge in Eq.~\eqref{eq:monopole} counts how many ``magnetic'' fluxes penetrate the surface $S$. A PHS pair of Weyl points, e.g., ${\bm k}_0$ and $-{\bm k}_0$, possesses the monopole charge $q _{\rm m}\!=\! + 1$ and $-1$, respectively. This indicates that the nodal points are topologically protected and a source of fictitious magnetic field ${\bm \Omega}_-({\bm k})$ in the ${\bm k}$ space. 

\begin{figure}[t]
\includegraphics[width=80mm]{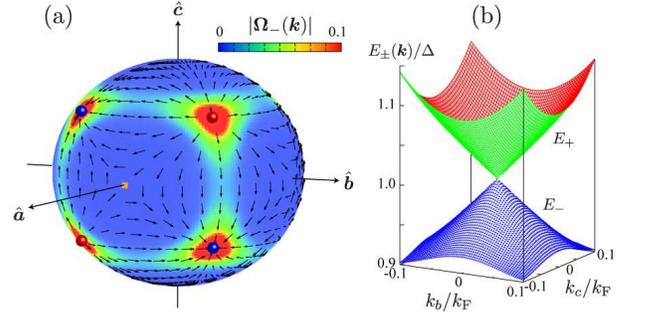}
\caption{(Color online) (a) Profiles of Berry curvature on the Fermi surface, ${\bm \Omega}_{n=-}({\bm k})$, constructed from $E_-({\bm k})$, where the color map shows the amplitude, $|\Omega _-({\bm k}_{\rm F})|$. The Berry curvature diverges at the level crossing lines ($\hat{\bm k}\parallel\hat{\bm a}$, $\hat{\bm b}$, or $\hat{\bm c}$). (d) Quasiparticle spectra, $E_{\pm}({\bm k})$, in the vicinity of the level crossing lines.
}
\label{fig:berry}
\end{figure}

In Fig.~\ref{fig:berry}(a), we plot the Berry curvature ${\bm \Omega}_-({\bm k})$ on the Fermi sphere ${\bm k}={\bm k}_{\rm F}$ for the cyclic $p$-wave state. The ``magnetic'' fluxes are generated by the Weyl points with $q_{\rm m}=+1$ and absorbed by the $q_{\rm m}=-1$ nodal points. It turns out that the nontrivial configuration of ${\bm \Omega}_-({\bm k})$ is also a source of the formation of topological Fermi arcs on the surface.

Let us now consider the low-energy quasiparticle structure of the cyclic state around the point nodes. We first show that the effective low-energy Hamiltonian for the cyclic state is described by the Weyl Hamiltonian. It is convenient to introduce a new Cartesian triad $(\hat{\bm n}_1,\hat{\bm n}_2,\hat{\bm n}_3)$, where $\hat{\bm n}_3 \!\equiv\! \hat{\bm n}_1\times \hat{\bm n}_2$ denotes one of the nodal directions, e.g., ${\bm k}_{0}$. In these basis, the low-energy part of the $4\times 4$ BdG Hamiltonian for the cyclic $p$-wave ($E_u$) state is decomposed into a pair of $2\times 2$ matrix as 
\beq
\mathcal{H}({\bm k}) \approx 
\mathcal{H}_+({\bm k}) 
\oplus
\mathcal{H}_-({\bm k}) .
\eeq
The $2\times 2$ submatrices are given by 
$\mathcal{H}_+({\bm k}) = \varepsilon _0 ({\bm k})\tau _3 + \sqrt{2}\hat{k}_1 \bar{\Delta}\tau _1$ and
$\mathcal{H}_-({\bm k}) = \varepsilon _0 ({\bm k})\tau _3 + \bar{\Delta}\hat{k}_1 \tau _1 + \bar{\Delta} \hat{k}_2 \tau _2$,
where ${\bm k}\!=\! k_1\hat{\bm n}_1 + k_2\hat{\bm n}_2 +k_3\hat{\bm n}_3$ and $\bar{\Delta}\equiv\Delta\omega^2$.
The former submatrix represents the fully gapped $E_+$ band, and the gap function is reduced to the polar state ($\bar{\Delta}\hat{k}_1$) around a Weyl point. In the vicinity of a Weyl point, the lower band with $E_-$ is described by the effective Hamiltonian, $\mathcal{H}_-({\bm k}) $, for the chiral $p$-wave state.
The mixing of $\mathcal{H}_+({\bm k})$ and $\mathcal{H}_-({\bm k})$ appears in the order of $k^2$. Hence, the low-energy structure of the cyclic $p$-wave state around a pair of nodes $q{\bm k}_0$ ($q=\pm 1$) is given by the Hamiltonian for Weyl-type Bogoliubov quasiparticles with a single pseudospin species, 
\beq
\mathcal{H}_-({\bm k}) = e^{\mu} _av^{a}_b\tau ^{b}(k_{\mu}-qk_{0,\mu}),
\label{eq:weyl}
\eeq
where the vielbein ${e}^{\mu}_a$ is defined as $(e^{\mu}_1,e^{\mu}_2,e^{\mu}_3) = (\hat{n}_{1,\mu},\hat{n}_{2,\mu},\hat{n}_{3,\mu})$ with the velocity tensor $v^{a}_b={\rm diag}(\bar{\Delta}/k_{\rm F},\bar{\Delta}/k_{\rm F},qv_{\rm F})$. This describes Weyl fermions with an effective electric charge $q$ coupled to the effective gauge field ${\bm k}_0$. 

The low-energy quasiparticles in the cyclic $d$-wave ($E_g$) state are also described by the Weyl-type Hamiltonian similar to Eq.~\eqref{eq:weyl} and the point nodes are identified as Weyl points with $q_{\rm m}=\pm 1$. The gap function is displayed in Fig.~\ref{fig:cyclic}(a).

We here mention that owing to the PHS in Eq.~\eqref{eq:phs}, a pair of Weyl fermions at ${\bm k}_{\rm node}$ and $-{\bm k}_{\rm node}$ behaves as three-dimensional Majorana fermions. To clarify this, we introduce the coordinates centered on the Weyl point, ${\bm K}\!\equiv\! {\bm k}-{\bm k}_{\rm node}$. Then, the four-component real quantum field, 
\beq
\psi ({\bm r})= \mathcal{C}\psi ({\bm r}), 
\eeq
can be constructed from a PHS pair of the single-species Weyl fermions as $\psi _{\alpha} ({\bm r})\!\equiv \! \sum _{\bm K}e^{i{\bm K}\cdot{\bm r}}\psi _{\alpha} ({\bm K})$ with $[c_{\alpha}({\bm K}), c_{\alpha}({\bm K}), c^{\dag}_{\alpha}(-{\bm K}), c^{\dag}_{\alpha}(-{\bm K})]^{\rm T}$. The low-energy Hamiltonian can be then recast into the Majorana-type Hamiltonian 
\beq
S_- = \int d^4x \bar{\psi}(x)\left(
i\partial _t - ie^{\mu} _av^{a}_b \gamma^b\partial _{\mu}
\right)\psi(x)
\eeq
where we have introduced  $(\gamma ^1,\gamma ^2,\gamma ^3) \!=\! (\mu _1\tau _1,\mu _1\tau _2,\mu _3)$ and $\bar{\psi}\!=\!(\tau _1 \psi)^{\rm T}$ with the Pauli matrices $\mu _{i}$ labeled by $q=\pm 1$. 
Hence, the low-energy structure of the cyclic phase is reduced to three-dimensional massless Majorana fermions. The Majorana fermion possesses pseudospin $1/2$ associated with the pairwise Weyl points and forms a quartet ($\psi _1,\cdots,\psi _4$) as a consequence of the tetrahedral point group symmetry.


It is seen in Fig.~\ref{fig:berry}(a) that the Berry fluxes form the quadrupole field around $\hat{\bm a}$, $\hat{\bm b}$, and $\hat{\bm c}$ axes. The center of the quadrupole field corresponds to the singularity in ${\bm \Omega}_-({\bm k})$. This singularity is attributed to the fact that the lower branch touches the upper energy branch, $E_+({\bm k})=E_-({\bm k})$, at ${\bm k}\!\parallel\! \hat{\bm a}$, ${\bm k}\!\parallel\! \hat{\bm b}$, and ${\bm k}\!\parallel\! \hat{\bm c}$ (see Fig.~\ref{fig:berry}(b)). We notice that although Weyl points are the sources of a nontrivial Chern number in the ${\bm k}$ space, it is ill-defined on the plane which intersects the level-crossing lines.

\section{Symmetry and topology of surface Majorana arcs}
\label{sec:arc}

A manifestation of nontrivial topological structure in nodal superconductors and superfluids is the appearance of surface Fermi arcs. Using the quasiclassical theory, we here show the evolution of surface Majorana arcs in the cyclic $p$-wave state with respect to the change of the surface orientation angles. 

In a typical Weyl superconductor such as chiral $p+ip$ state, the surface Fermi arc connecting the projections of the bulk Weyl points is protected by the first Chern number. In contrast, the one-dimensional winding number associated with a chiral symmetry is responsible for the existence of surface Fermi arc in time-reversal-invariant superconductors/superfluids.~\cite{satoPRB11,tsutsumiJPSJ13,mizushimaJPCM15,sasaki15,mizushimaJPSJ16} We also notice that in chiral superconductors with a line node the fragileness of surface Fermi arc was discussed in terms of a one-dimensional winding number associated with the pseudo-time-reversal symmetry.~\cite{kobayashiPRB15} For the nonunitary cyclic state which can be regarded as an admixture of full gap and point node gap, however, we will demonstrate below that the surface Majorana arcs are protected by two different topological invariants: the first Chern number ${\rm Ch}_1$ and one-dimensional winding number $w_{\rm 1d}$.

\begin{figure}[t]
\includegraphics[width=80mm]{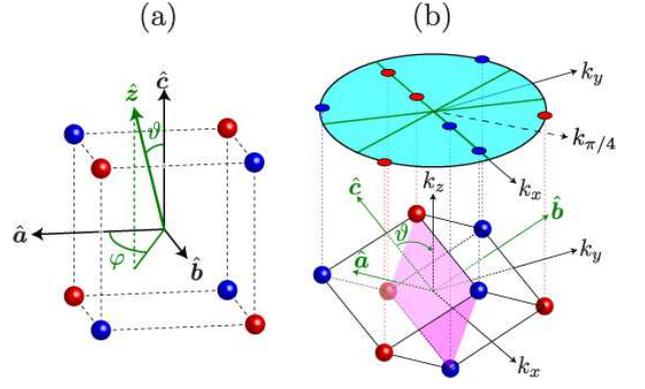}
\caption{(Color online) (a) Surface orientation with respect to the cyclic order parameter under cubic crystalline symmetry. The surface normal axis, $\hat{\bm z}$, is parameterized with $\varphi$ and $\vartheta$, where $\vartheta$ denotes the relative angle between $\hat{\bm z}$ and $\hat{\bm c}$. (b) Weyl points projected onto the surface momentum space ${k}_x$-${k}_y$ for $\varphi = \pi/4$ and $\vartheta/\pi = 0.4$. The shaded $k_x$-$k_z$ plane shows the $P_2$ symmetric momentum plane.
}
\label{fig:surface}
\end{figure}

\subsection{Evolution of surface Majorana arcs}

To calculate the surface density of states in nonunitary superconductors, we here utilize the quasiclassical theory. The central object of the quasiclassical theory is the propagator, $g(\hat{\bm k},{\bm r};\varepsilon _n)$, that contains both quasiparticles and superfluidity in equal footing. The propagator is obtained from the Matsubara Green's function ${G}({\bm k},{\bm r};\varepsilon _n)$ by integrating ${G}$ over a shell $v_{\rm F}|k-k_{\rm F}| < E_{\rm c} \ll E_{\rm F}$,~\cite{serene}
${g}(\hat{\bm k},{\bm r};\varepsilon _n) \!=\! \frac{1}{a} \int^{+E_{\rm c}}_{-E_{\rm c}} d\xi _{\bm k}
{\tau}_z{G}({\bm k},{\bm r};\varepsilon _n)$. The normalization constant $a$ corresponds to the weight of the quasiparticle pole in the spectral function. The quasiclassical propagator $\underline{g}$ that is a $4\times 4$ matrix in particle-hole and spin spaces is parameterized with spin Pauli matrices $\sigma _{\mu}$ as
\beq
\underline{g} = \left(
\begin{array}{cc}
g_{0} + {\sigma}_{\mu} g_{\mu} & i\sigma _y f_0 + i {\sigma}_{\mu} {\sigma}_y f_{\mu}  \\ 
i\sigma _y \bar{f}_0 +i \sigma _y {\sigma}_{\mu}\bar{f}_{\mu}  & \bar{g}_{0} + {\sigma}^{\rm T}_{\mu} \bar{g}_{\mu}
\end{array}
\right).
\label{eq:g}
\eeq
The off-diagonal propagators are composed of spin-singlet and triplet Cooper pair amplitudes, $f_0$ and $f_{\mu}$. 

The quasiclassical propagator is governed by the transport-like equation \eqref{eq:eilen} supplemented by the normalization condition in Eq.~\eqref{eq:norm}. We here consider a semi-infinite system, $z\in [0,\infty)$, having a specular surface at ${\bm r}={\bm r}_{\rm surf} \!=\! (x,y,0)$, where $z$ denotes the distance from the surface. A quasiparticle incoming to the surface along the trajectory of ${\bm k}$ is specularly scattered by the wall to the quasiparticle state with 
$\underline{\bm k}= {\bm k} - 2\hat{\bm z}(\hat{\bm z}\cdot{\bm k})$.
The specular boundary condition is imposed on the quasiclassical propagator as 
\beq
g(\underline{\hat{\bm k}},{\bm r}_{\rm surf};\varepsilon _n)
= g(\hat{\bm k},{\bm r}_{\rm surf};\varepsilon _n).
\eeq
The further details on the formalism and the numerical procedure are described in Appendix A.

As shown in Fig.~\ref{fig:surface}(a), we parameterize the surface orientation ($\hat{\bm z}$) with $(\varphi,\vartheta)$ relative to the crystal coordinates. It is now convenient to introduce new coordinates, ($\hat{\bm x},\hat{\bm y},\hat{\bm z}$), where the crystal coordinates, $(\hat{\bm a},\hat{\bm b},\hat{\bm c})$, are obtained by rotating ($\hat{\bm x},\hat{\bm y},\hat{\bm z}$) with $R\equiv R_b(-\vartheta)R_z(-\varphi)$, as $(\hat{ a}_{\mu},\hat{b}_{\mu},\hat{c}_{\mu})={R}_{\mu \nu}(\hat{x}_{\nu},\hat{y}_{\nu},\hat{z}_{\nu})$, where $R_n(\theta)$ stands for the rotation matrix by the angle $\theta$ about the $n$ axis. The relative rotation of the crystal coordinates from the surface orientation transforms the  order parameter tensor to $A_{\mu i} = R_{\mu \nu}\tilde{A}_{\nu j}R_{ij}$, where $\tilde{A}_{\nu j}$ is the order parameter tensor for $(\hat{\bm a},\hat{\bm b},\hat{\bm c})=(\hat{\bm x},\hat{\bm y},\hat{\bm z})$.


The schematic picture on the configuration of monopole-antimonopole pairs and surface momentum space for $(\varphi/\pi,\vartheta/\pi) =(0.25,0.4)$ is depicted in Fig.~\ref{fig:surface}(b). The surface configuration with $\varphi/\pi = 0.25$ preserves the $P_2$ symmetry that is the the order-two discrete symmetry introduced in Eq.~\eqref{eq:p2}. We will show below that some of the Fermi arcs are protected by the one-dimensional winding number associated with the $P_2$ symmetry but not the Chern number. 

To clarify the structure of surface Majorana arcs and their topological and symmetry backgrounds, we first show the distribution of the zero energy quasiparticle states in the surface momentum space. We start to introduce the angle-resolved surface density of states
\beq
{N}_{\rm S}({k}_x,{k}_y,E) \equiv \sum _{{\rm sgn}(\hat{k}_z)}{N}(\hat{\bm k},z=0,E).
\label{eq:sdosk}
\eeq
The surface Brillouin zone is represented by $({k}_x,{k}_y)$. The ${\bm k}$-resolved local density of states, ${N}(\hat{\bm k},{\bm r};E)$, is obtained from Eq.~\eqref{eq:eilen} with the analytic continuation $i\varepsilon _n \rightarrow E+i0_+$ in the diagonal part of the quasiclassical propagator
\beq
{N}(\hat{\bm k},{\bm r};E) = -\frac{{N}_{\rm F}}{\pi}{\rm Im}g_0(\hat{\bm k},{\bm r};\varepsilon _n \rightarrow -iE+0_+),
\label{eq:dosk}
\eeq
where ${N}_{\rm F} \!=\! \int \frac{d\hat{\bm k}}{(2\pi)^3|{\bm v}_{\rm F}(\hat{\bm k})|}$ is the total density of states at the Fermi surface in the normal state. In Eq.~\eqref{eq:sdosk}, $\sum _{{\rm sgn}(\hat{k}_z)}$ denotes the sum over $\hat{k}_z$ which satisfies $\hat{k}_z= \pm \sqrt{1-\hat{k}^2_x-\hat{k}^2_y}$, and $({k}_x,{k}_y)$ is a set of the surface momenta.

\begin{figure}[t]
\includegraphics[width=80mm]{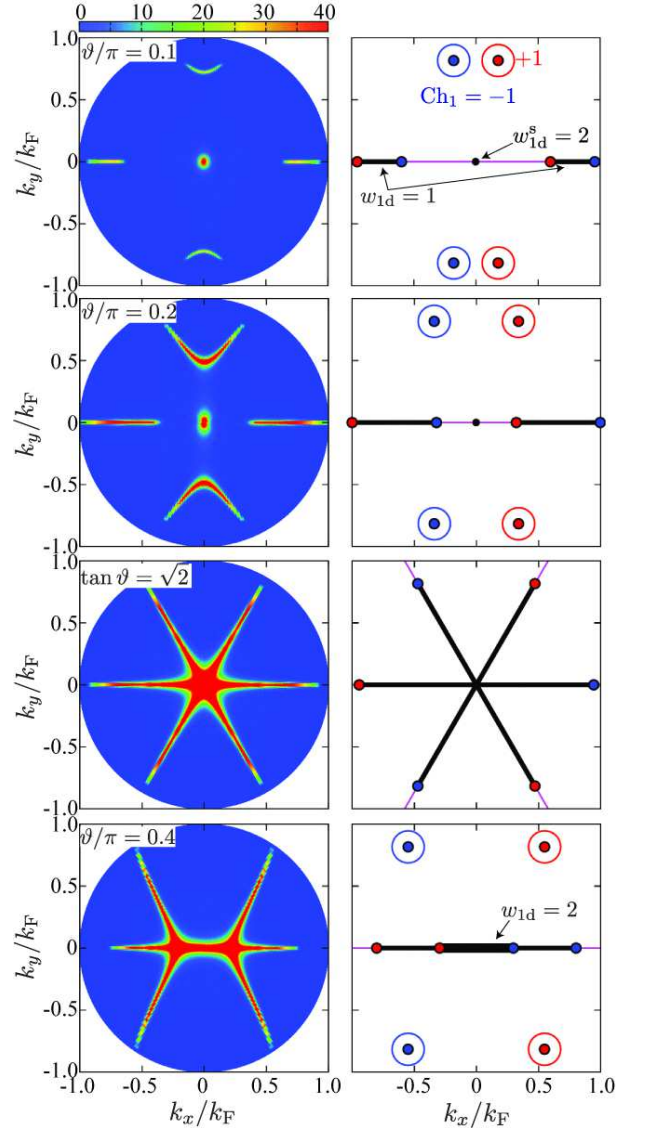}
\caption{(Color online) (Left) Angle-resolved zero-energy density of states on the surface, ${N}_{\rm S}({k}_x,{k}_y,E=0)$, in the cyclic $p$-wave state for various surface orientation angles, $(\varphi/\pi,\vartheta/\pi) = (0.25,0.1)$, $(0.25,0.2)$, $(0.25,0.4)$, $(0.25,\vartheta _{111})$, and $(0.25,0.4)$. The set of angles, $\varphi = \pi/2$ and $\tan\vartheta _{111}={\sqrt{2}}$, correspond to the $[111]$ surface. (Right) Projected Weyl points and topological invariants relevant to Fermi arcs, the first Chern number ${\rm Ch}_1\!=\!\pm 1$ and one-dimensional winding number $w_{\rm 1d}({k}_x,{k}_y=0)$, in the surface momentum space. The thin (green) lines denote the $P_2$ symmetric plane in which $w_{\rm 1d}\in \mathbb{Z}$ is well-defined.}
\label{fig:ldosk}
\end{figure}

Figure \ref{fig:ldosk} shows the angle-resolved zero-energy density of states on the surface Brillouin zone, ${N}_{\rm S}({k}_x,{k}_y,E=0)$, in the cyclic $p$-wave state for various surface orientation angles, $(\varphi/\pi,\vartheta/\pi) = (0.25,0.1)$, $(0.25,0.2)$, $(0.25,0.4)$, $(0.25,\vartheta _{111})$, and $(0.25,0.4)$, where $\vartheta _{111}=\tan^{-1}(\sqrt{2}) \approx 0.6(\pi/2)$ stands for the orientation angle for the [111] surface. In the right panels of Fig.~\ref{fig:ldosk}, we plot the bulk Weyl points projected onto the surface ${\bm k}$-space. These Weyl points with monopole charges $q_{\rm m} = \pm 1$ are sources of a nontrivial Chern number ${\rm Ch}_1$, which characterizes the topological structure of the surface Majorana arcs. As mentioned in Sec.~\ref{sec:weyl}, the band crossing at the $\hat{\bm a}$, $\hat{\bm b}$, and $\hat{\bm c}$ axes gives rise to the singularity in the Berry curvature in the bulk Brillouin zone. Since this prevents an well-defined Chern number in a two-dimensional plance that contains the singularities, we define the Chern number ${\rm Ch}_1$ on a small surface enclosing each Weyl point. As the panels show, the surface Majorana arcs connect the projections of the bulk Weyl points.




\subsection{Combined symmetry and topological invariant}

We have seen that owing to the Weyl points in $E_-({\bm k})$, the cyclic state possesses a nontrivial Berry curvature in the momentum space. For unitary states with a single pair of Weyl points, the Fermi arc appears as a consequence of ${\rm Ch}_1\!\neq\!0$ well-defined in a sliced two-dimensional momentum plane. We here introduce another type of topological invariants in connection with crystalline symmetries. Both the Chern number and winding number are indispensable for understanding the structure of surface Majorana arcs in cyclic $p$-wave states.

To introduce the winding number, we first clarify the discrete symmetry of the cyclic $p$-wave state. We here fix $\varphi$ to be $\pi/4$ which is the most symmetric surface configuration. Although cubic crystals with the $O_h$ symmetry possess mirror refection planes associated with the $C_2$ axes, the formation of the cyclic pairing spontaneously breaks the crystalline symmetry into the tetrahedral symmetry and $[110]$ mirror reflection planes disappear. Hence, the cyclic $p$-wave state in Eq.~\eqref{eq:dcyclic} spontaneously breaks $\mathcal{T}$ and $M$, independently. However, it remains invariant under the combined symmetry, 
\beq
\mathcal{T}M\Delta ({\bm k})(\mathcal{T}M)^{-1} =-\Delta (-k_x,k_y,-k_z).
\label{eq:p2d}
\eeq
The mirror operator $M \equiv -i{\bm \sigma}\cdot\hat{\bm n}$ flips the momentum and spin as ${\bm \sigma}\!\mapsto -{\bm \sigma}+2\hat{\bm n}({\bm \sigma}\cdot\hat{\bm n})$ and ${\bm k}\!\mapsto {\bm \sigma}-2\hat{\bm n}({\bm k}\cdot\hat{\bm n})$, respectively, where $\hat{\bm n}$ denotes the [110] surface orientation. Equation \eqref{eq:p2d} implies that the cyclic state is invariant under the combined discrete symmetry, ${P}_2 =\mathcal{T}\mathcal{M}$,
\beq
P_2 \mathcal{H}({\bm k})P^{-1}_2 = \mathcal{H}(-k_x,k_y,-k_z),
\label{eq:p2}
\eeq
where $\mathcal{M}=M \oplus (\sigma _yM\sigma _y)$ stands for the mirror operator in the Nambu space.

Combining it with the PHS in Eq.~\eqref{eq:phs}, one obtains the chiral symmetry, $\Gamma \equiv -i\mathcal{C}{P}_2$, satisfying $\{\Gamma, \mathcal{H}(k_x,0,k_z)\}\!=\! 0$. Let $U$ be a unitary matrix which diagonalizes $\Gamma$ as $U\Gamma U^{\dag}= {\rm diag}(+1,+1,-1,-1)$. Then, $U$ transforms the BdG Hamiltonian to the off-diagonal form
\beq
U\mathcal{H}(k_x,0,k_z)U^{\dag} = \left( 
\begin{array}{cc}
0 & q(k_x,k_z) \\ q^{\dag}(k_x,k_z) & 0
\end{array}
\right).
\eeq 
As long as the symmetry is maintained, the one dimensional winding number in the chiral symmetric momenta ${\bm k}\!=\!(k_x,0,k_z)$ is defined as the topological invariant relevant to the surface Fermi arc at $k_y\!=\!0$,~\cite{satoPRB11}
\begin{align}
w_{\rm 1d}(k_x) &= -\frac{1}{4\pi i}\int^{+\pi}_{-\pi} dk_z{\rm tr}
[ 
\Gamma \mathcal{H}^{-1}({\bm k})\partial _{k_z} \mathcal{H}({\bm k})
]_{k_y = 0} \nn \\
&= \frac{1}{2\pi}{\rm Im}\int^{+\pi}_{-\pi} dk_z
\partial _{k_z}\ln \det q(k_x,k_z)
\label{eq:w1d}
\end{align}
For the cyclic state with the orientation angle $\vartheta$, the determinant of the $q$-matrix is given as
\begin{align}
\det q(k_x,k_z) =& -\left[\varepsilon(k_x,k_z)\right]^2
- \frac{\Delta^2}{k^2_{\rm F}}\left(1-\frac{3}{2}\cos^2\!\vartheta\right)k^2_x \nn \\
&- \frac{\Delta^2}{k^2_{\rm F}}\left(1-\frac{3}{2}\sin^2\!\vartheta\right)k^2_z
+ \frac{3}{2}\frac{\Delta^2}{k^2_{\rm F}}\sin(2\vartheta)k_xk_z \nn \\
& + i\sqrt{3}\frac{\Delta}{k_{\rm F}}\varepsilon(k_x,k_z)\left( \cos\vartheta k_x + \sin\vartheta k_z\right).
\end{align}
The chiral symmetry guarantees that all energy eigenstates are labeled by the eigenstates $\Gamma \!=\! \pm 1$. $w_{\rm 1d}$ is identical to the difference in the number of zero-energy states in each chiral subsector, $|w_{\rm 1d}| \!=\! |N_+-N_-|$.~\cite{satoPRB11} In contrast to ${\rm Ch}_1$, the Fermi arc is only protected by the $P_2$ symmetry. 

To evaluate Eq.~\eqref{eq:w1d}, it is convenient to introduce the following two-dimensional unit vector, $\hat{\bm m}=(\hat{m}_1,\hat{m}_2)$, where $\hat{m}_1(k_x,k_z)\equiv {\rm Re}\det q(k_x,k_z)/|\det q(k_x,k_z)|$ and $\hat{m}_2(k_x,k_z)\equiv {\rm Im}\det q(k_x,k_z)/|\det q(k_x,k_z)|$. Then, Eq.~\eqref{eq:w1d} can be recast into 
\beq
w_{\rm 1d}(k_x) = \frac{1}{2\pi} \int^{+\pi}_{-\pi}dk_z \epsilon^{ij}\hat{m}_i(k_x,k_z) \partial _{k_z} \hat{m}_j(k_x,k_z),
\eeq
which counts how many the one-dimensional momentum along $k_z$ wraps the target space represented by $\det q/|\det q |\in S^1$ for a fixed $k_x$  ($i,j = 1$ and $2$). Since only the neighborhood of the zeros of $\varepsilon (k_x,0,k_z)$ contributes to the integral, the winding number is simplified to the sum at $k_0$ that satisfies $\varepsilon(k_x,0,k_0)=0$ as
\begin{align}
w_{\rm 1d}(k_x) =& \frac{1}{2} \sum _{k_0\in {\rm F.S.}} {\rm sgn}[\partial _{k_z}\hat{m}_2(k_x,k_0)] 
\left\{
1 + {\rm sgn}[ \hat{m}_1(k_x,k_0)]
\right\}.
\end{align} 
For $k_x=0$, this can be evaluated as 
\beq
w_{\rm 1d}(0) = \left\{
\begin{array}{ll}
0 & \mbox{ for $\vartheta < \vartheta _{111}$}\\
\\
-2 & \mbox{ for $\vartheta > \vartheta _{111}$}
\end{array}
\right. ,
\eeq
when ${\bm H}\cdot\hat{y}=0$.

\begin{figure}[t]
\includegraphics[width=60mm]{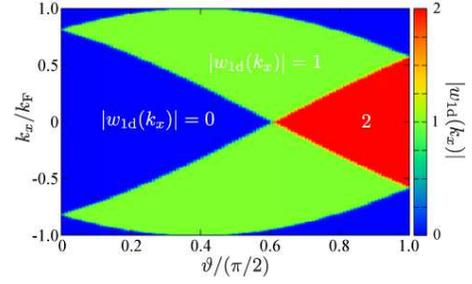}
\caption{Evolution of the one-dimensional winding number, $w_{\rm 1d}({k}_x,{k}_y=0)$, defined in Eq.~\eqref{eq:w1d}, where we fix $\varphi = 0.25$. 
}
\label{fig:w1d}
\end{figure}

Figure \ref{fig:w1d} shows the evolution of the one-dimensional winding number, $w_{\rm 1d}({k}_x)$ with respect to the surface orientation angle $\vartheta$. Here we fix $\varphi = 0.25$ so that the $P_2$ symmetry is preserved even in the presence of the surface. It is seen that the winding number becomes nontrivial, $|w_{\rm 1d}(k_x)|=1$, in the segment connecting the PHS pair of Weyl points, which implies the $P_2$ symmetry protection of the surface Majorana arc along the ${k}_x$ axis. In the cases of $\varphi/\pi = 0.1$ and $0.2$ in Fig.~\ref{fig:ldosk}, therefore, two Majorana arcs on the ${k}_x$ axis can be protected by both the one-dimensional winding number $|w_{\rm 1d}(k_x)|=1$ and the first Chern number. The former is well-defined unless the $P_2$ symmetry is broken, while the latter is robust regardless of the $P_2$ symmetry breaking.

At $\vartheta _{111}$ and $\varphi = 0.25\pi$, the $\omega C_3$ rotation symmetry about a normal surface is maintained even in the presence of the surface. In this configuration, the surface maintains the three $P_2$ symmetric plane and all bulk Weyl points are placed on the $P_2$ symmetric planes. This indicates that the three Majorana arcs in Fig.~\ref{fig:ldosk} ($\tan\vartheta=\sqrt{2}$) originate from the $P_2$ symmetry protected winding number as well as ${\rm Ch}_1$. For $\vartheta>\vartheta _{111}$, as shown in Fig.~\ref{fig:w1d}, the topological invariant takes $|w_{\rm 1d}(k_x)|=2$ in the central region of the ${k}_x$ axis and $|w_{1d}(k_x)|=1$ otherwise. For $\varphi = 0.4\pi$ in Fig.~\ref{fig:ldosk}, therefore, the central region of the Fermi arc on $k_y=0$ is protected solely by $|w_{\rm 1d}(k_x)|=2$, while the outer arcs are characterized by both $|w_{\rm 1d}|=1$ and $|{\rm Ch}_1|=1$. 

\begin{figure}[t]
\includegraphics[width=85mm]{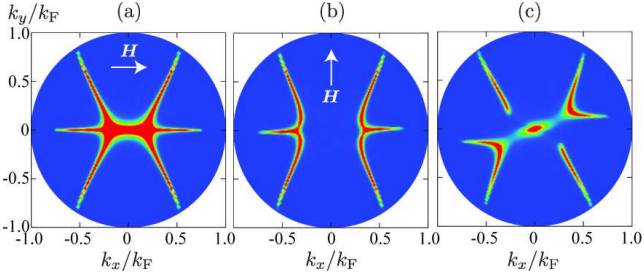}
\caption{Momentum resolved zero-energy density of states, ${N}_{\rm S}({k}_x,{k}_y,E=0)$, in the cyclic $p$-wave state for $(\varphi/\pi,\vartheta/\pi) = (0.25,0.4)$ (a,b) and (0.15,0.4) (c). The applied magnetic field in (a) preserves the $P_2$ symmetry, while it breaks the symmetry in (b). }
\label{fig:jiba}
\end{figure}

In addition to $w_{\rm 1d}$, we can introduce another winding number that ensures the existence of zero energy states at ${k}_x={k}_y = 0$. It is obvious that for $\vartheta = 0$, the gap function of the cyclic $p$-wave state with $k_x = k_y = 0$ reduces to that of the polar phase, ${\bm d}(0,0,k_z) = \Delta \hat{k}_z\hat{\bm z}$, at which the TRS emerges. The emergent TRS at $k_x=k_y =0$ leads to the chiral symmetry as a combination of the TRS and PHS. 

For $\varphi = \pi/4$, the BdG Hamiltonian has the accidental symmetry, which is called the pseudo TRS.~\cite{kobayashiPRB15} At $k_x=k_y = 0$, the cyclic order parameter is given as $\Delta (0,0,k_z) = \Delta [\sqrt{\frac{3}{2}}\sin\vartheta+\frac{3}{2\sqrt{2}}\sin (2\vartheta) \sigma _z +\sqrt{2}(1-\frac{3}{2}\sin^2\vartheta)\sigma _x]k_z$. The $2\times 2$ matrix is diagonalized to $V\Delta (0,0,k_z)V^{\rm t}={\rm diag}(ak_z,bk_z)$, where $V\equiv \cos\frac{\phi _0 }{2} -i\sigma _y \sin \frac{\phi _0}{2}$ is an ${\rm SU}(2)$ matrix representing the spin rotation about the $\hat{\bm y}$ axis and $a, b \in \mathbb{R}$ stand for the polar-like gap amplitudes. The rotation angle, $\phi _0$, is taken so as to satisfy $\sqrt{2}(1-\frac{3}{2}\sin^2\vartheta)\cos\phi _0 + \sqrt{\frac{3}{2}}\sin\vartheta\sin \phi _0 = 0$. Hence, the BdG Hamiltonian for $k_x=k_y=0$ has the following pseudo TRS
\beq
\mathcal{V}^{\dag}\mathcal{T} \mathcal{V} \mathcal{H}(0,0,k_z)\mathcal{V}^{\dag}\mathcal{T}^{-1} \mathcal{V} 
= \mathcal{H}(0,0,-k_z),
\eeq
where $ \mathcal{V} = V \oplus V^{\ast}$ is the spin rotation matrix in the Nambu space. Using the spin rotation operator, the chiral operator is defined as $\Gamma^{\rm s} = i\mathcal{V}^{\dag}\mathcal{CT} \mathcal{V} $. With the chiral operator, we define the one-dimensional winding number as
\beq
w^{\rm s}_{\rm 1d} = -\frac{1}{4\pi i}\int^{+\pi}_{-\pi} dk_z {\rm tr}\left[
\Gamma^{\rm s}\mathcal{H}^{-1}(0,0,k_z)\partial _{k_z} \mathcal{H}(0,0,k_z)
\right].
\eeq
The winding number is estimated as 
\beq
w^{\rm s}_{\rm 1d} = \sum _{k_0 \in {\rm F.S.}} {\rm sgn}[ \partial _{k_z}\varepsilon (0,0,k_z) ]{\rm sgn}(k_z).
\label{eq:w1ds}
\eeq
For the case of a spherical Fermi surface, it yields $w^{\rm s}_{\rm 1d} = 2$ which ensures the existence of the zero energy states at $k_x=k_y=0$ as shown in Fig.~\ref{fig:ldosk}. Since the chiral symmetry originates in the accidental spin rotation symmetry, the zero energy states at $k_x=k_y=0$ will be sensitive to a perturbation with broken spin rotation symmetry, {\it e.g.}, spin-orbit interactions.

To demonstrate the fragileness of the Majorana arcs with $|w_{\rm 1d}|=2$, in Fig.~\ref{fig:jiba} we display the field-orientation- and $\varphi$-dependence of the surface Majorana arcs in the cyclic $p$-wave state. We notice that a magnetic field along the [110] mirror reflection plane maintains the $P_2$ symmetry because the mirror reflection of the field, ${\bm H}\mapsto-{\bm H}$, can be compensated by the TRS. When the applied field is misoriented from the [110] mirror plane, it explicitly breaks the $P_2$ symmetry. It is seen from Fig.~\ref{fig:jiba}(a) that the $P_2$ symmetric field does not alter the structure of surface Majorana arcs, while the $|w_{\rm 1d}|=2$ segment of the surface Majorana arcs disappears in the presence of the $P_2$ symmetry breaking field (Fig.~\ref{fig:jiba}(b)). This implies that the surface Majorana arc with $|w_{\rm 1d}|=2$ is not characterized by the Chern number and protected by solely the $P_2$ symmetry. The fragileness of the Majorana arc with $|w_{\rm 1d}|=2$ is attributed to the Ising-like anisotropy of the zero energy states solely protected by $w_{\rm 1d}$.~\cite{satoPRB2009,mizushimaPRL12,mizushimaNJP13,tsutsumiJPSJ13,mizushimaJPCM15,mizushimaJPSJ16,shiozakiPRB14,xiong17}


\begin{figure}[t]
\includegraphics[width=80mm]{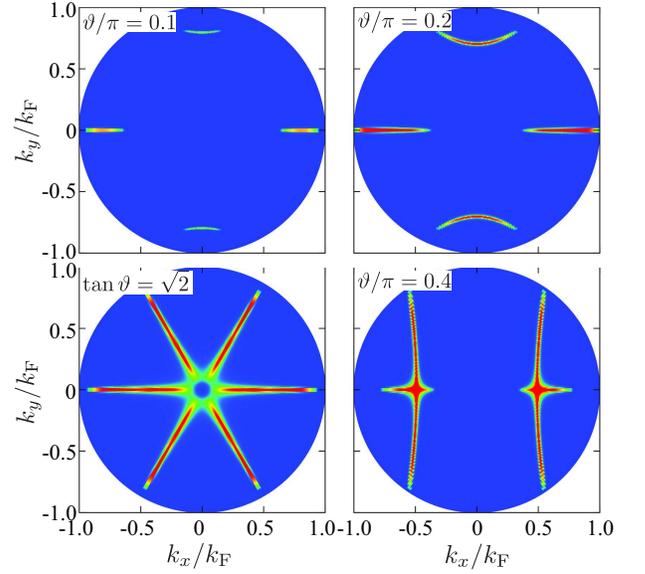}
\caption{${N}_{\rm S}({k}_x,{k}_y,E=0)$ in the cyclic $d$-wave state: $(\varphi/\pi,\vartheta/\pi) = (0.25,0.1)$, $(0.25,0.2)$, $(0.25,\vartheta _{111})$, and $(0.25,0.4)$.}
\label{fig:Eg}
\end{figure}

For comparison, in Fig.~\ref{fig:Eg}, we show the momentum resolved surface density of states, ${N}_{\rm S}({k}_x,{k}_y,E=0)$ in the cyclic $d$-wave state with the gap function in Eq.~\eqref{eq:dcyclic}. The Fermi arc structure was discussed in Ref.~\onlinecite{ishikawaJPSJ13} in terms of the Andreev bound states with the $\pi$-phase shift. In contrast to Fig.~\ref{fig:ldosk}, the Fermi arc characterized by $|w_{\rm 1d}|=2$ disappears in the cyclic $d$-wave case for $\vartheta > \vartheta _{111}$. All Fermi arcs connecting the Weyl points are characterized solely by a nontrivial Chern number.

\subsection{Van Hove singularities in the surface bound states}

\begin{figure}[t]
\includegraphics[width=85mm]{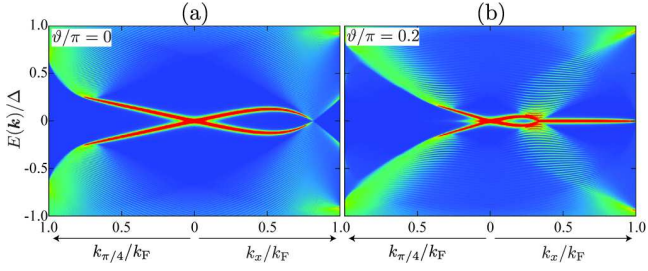}
\caption{Angle-resolved surface density of states in cyclic $p$-wave states for $(\varphi,\vartheta) = (\pi/4, 0)$ (a) and $(\pi/4,\pi/5)$ (b). $\bar{k}_x$ and $k_{\pi/4}$ stand for the nodal direction and the anti-nodal direction in the $P_2$ symmetric plane (see Fig.~\ref{fig:surface}(b)), respectively.}
\label{fig:sdos1}
\end{figure}

In Fig.~\ref{fig:sdos1}, we display the angle-resolved surface density of states, $N_{\rm S}(k_x,k_y,E)$ defined in Eq.~\eqref{eq:sdosk}, in cyclic $p$-wave states for different surface orientations. The gapless linear dispersion appears at $k_x=k_y=0$, which reflects the nontrivial topological invariant $w^{\rm s}_{\rm 1d}=2$ in Eq.~\eqref{eq:w1ds}. As mentioned above, the topological invariant is attributed to the pseudo TRS associated with a spin rotation. Hence, although the zero energy state survives for arbitrary misorientation angle $\vartheta$, it may be sensitive to perturbations with breaking the symmetry, such as, spin-orbit coupling. 

It is seen from Fig.~\ref{fig:sdos1} that there is the anisotropic gapless cone around $k_x=k_y=0$. The dispersion along the anti-nodal direction ($k_{\pi}/4$) is linear, while along the nodal direction ($k_x$) it is merged to the continuum states at the point node $(k_x/k_{\rm F},k_y/k_{\rm F}) = (\sqrt{2/3},0)$ and possesses the almost flat region at finite energies around $E/\Delta \approx 0.2$ in Fig.~\ref{fig:sdos1}(a). As the surface orientation angle $\vartheta$ is deviated from $\vartheta = 0$, the Majorana Fermi arcs develop along $k_x$ and the flat region disappears. The appearance of the flat region in the dispersion results in van Hove singularities in the surface density of states, while the topologically protected Majorana arcs lead to a sharp peak of the surface density of states at $E=0$. This implies the evolution of the surface density of states from the split peak structure at $E/\Delta \approx \pm 0.2$ to the single peak structure at $E=0$ when $\vartheta$ approaches $\vartheta _{111}$. 

\section{Tunneling conductance in U$_{1-x}$Th$_x$Be$_{13}$ superconducting junctions}
\label{sec:tunnel}

Lastly, using the Blonder-Tinkham-Klapwijk (BTK) theory,~\cite{btk} we calculate tunneling conductance spectra in U$_{1-x}$Th$_x$Be$_{13}$ superconducting junctions. For the gap symmetry of U$_{1-x}$Th$_x$Be$_{13}$, there are two competitive scenarios -- the degenerate $E_u$ scenario~\cite{shimizu17} and the accidental scenario.~\cite{sigristPRB89,sigrist} In the accidental scenario, the order parameter is constructed from two different representations of the $O_h$ symmetry. Based on the numerical calculation of tunneling conductance for both the scenarios, we discuss how the tunneling spectra capture a hallmark of topologically protected Majorana arcs in the nonunitary cyclic state.


The BTK theory was generalized to nonunitary superconductors.~\cite{asanoPRB03,bohloul} Following \onlinecite{asanoPRB03,bohloul}, we consider a junction system composed of a normal metal ($z<0$) and a superconductor ($z>0$), and the insulating interface at $z=0$ is modeled as a $\delta$-function potential of height $H$. We here consider standard scattering and transmission processes of electrons injected from the metal side.~\cite{kashiwayaRPP} We suppose that an electron is injected into the superconductor from the $-\hat{\bm z}$ direction with momentum ${\bm k}$ and spin $s = \uparrow,\downarrow$, where $k_{z}>0$. At the interface, the electron may be reflected as a hole with momentum $-{\bm k}$ or as an electron with $\underline{{\bm k}}= {\bm k} - 2 \hat{\bm z}({\bm k}\cdot\hat{\bm z})$. The former represents the Andreev reflection, while the latter is the normal reflection. The wavefunction for a spin-$s$ incident electron in the normal side is given by the four-component spinor as 
\beq
{\bm \psi}^{\rm N}_s({\bm r}) = e^{i{\bm k}\cdot{\bm r}}\left(
\begin{array}{c}
1 \\ 0 \\ a_{s\uparrow}(E) \\ a_{s\downarrow}(E) 
\end{array}
\right)
+ e^{i\underline{\bm k}\cdot{\bm r}}
\left(
\begin{array}{c}
b_{s\uparrow}(E) \\ b_{s\downarrow}(E) \\ 0 \\ 0
\end{array}
\right).
\eeq
The coefficients, $a$ and $b$, represent the reflection coefficients of the Andreev and normal reflections, respectively.

\begin{figure*}[t!]
\includegraphics[width=170mm]{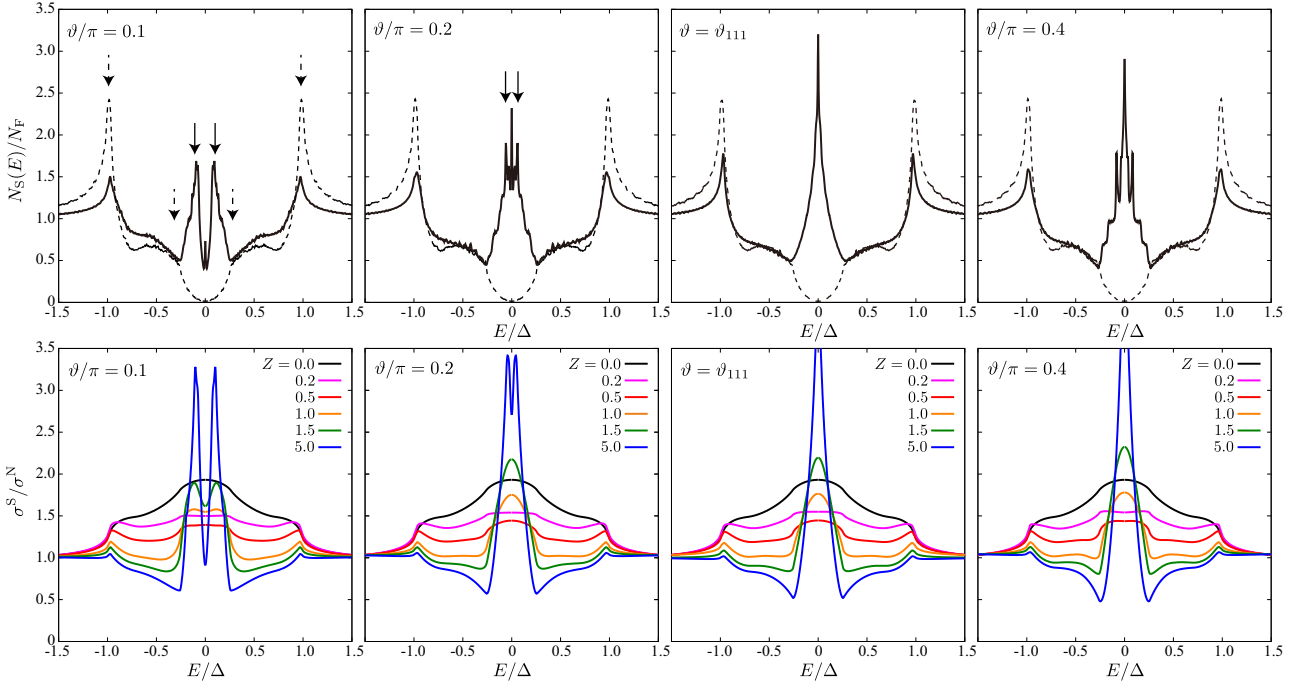}
\caption{(top) Surface density of states for the cyclic $p$-wave state for various orientation angles: $(\varphi/\pi,\vartheta/\pi) =(0.25,0.1)$, $(0.25,0.2)$, $(0.25,\vartheta _{111})$, and $(0.25,0.4)$. The solid (dashed) curves show the surface (bulk) density of states. (bottom) Normalized tunneling conductance $\sigma^{\rm S}/\sigma^{\rm N}$ for various the barrier potential $Z$.}
\label{fig:sdos2}
\end{figure*}

It may also be transmitted into the superconductor ($z>0$) as an ``electron-like'' quasiparticle with momentum ${\bm k}^{\prime}$ ($k^{\prime}_{z}>0$) or as a ``hole-like'' quasiparticle with $-\underline{{\bm k}}^{\prime} = -{\bm k} + 2 \hat{\bm z}({\bm k}^{\prime}\cdot\hat{\bm z})$. Continuity of the wavefunction at the interface requires $k_{x} = k^{\prime}_{x}$, $k_{y}=k^{\prime}_{y}$, and $k_{z}\sin\theta  = k^{\prime}_{z}\sin \theta^{\prime}$, where $\theta$ and $\theta^{\prime}$ are the polar angles on either side of the barrier. For simplicity, we will assume $k_{z} \approx k^{\prime}_{z}$. In the superconductor side, therefore, the general form of the wavefunction for transmitted quasiparticles is 
\begin{align}
{\bm \psi}^{\rm S}_s({\bm r}) =& e^{i{\bm k}\cdot{\bm r}}
\left(
c_{+}\varphi^{\rm p}_+ ({\bm k}) + c_{-}\varphi^{\rm p}_- ({\bm k}) 
\right) \nn \\
& + e^{-i\underline{\bm k}\cdot{\bm r}}
\left(
d_{+}\varphi^{\rm h}_+ (-{\bm k}) + d_{-}\varphi^{\rm h}_- (-{\bm k}) 
\right).
\label{eq:psis}
\end{align}
The coefficients, $c$ and $d$, represent the transmission coefficient of the electron-like quasiparticle and that of the hole-like quasiparticle, respectively. In superconducting states, the BdG Hamiltonian is diagonalized by using the Bogoliubov transformation matrix, $U({\bm k})\equiv [{\bm \varphi}^{\rm p}_+({\bm k}),{\bm \varphi}^{\rm p}_-({\bm k}), \mathcal{C}{\bm \varphi}^{\rm p}_+(-{\bm k}),\mathcal{C}{\bm \varphi}^{\rm p}_-(-{\bm k})]$, as 
\beq
U^{\dag}({\bm k})\mathcal{H}({\bm k})U({\bm k}) = \left( 
\begin{array}{cccc}
E_+ & & & \\
& E_- & &  \\
& & -E_+ & \\
& & & -E_-
\end{array}
\right).
\eeq
Therefore, the wavefunctions,
\beq
\left[{\bm \varphi}^{\rm p}_+,{\bm \varphi}^{\rm p}_-\right]
= \left(
\begin{array}{c}
\hat{u}({\bm k}) \\ \hat{v}^{\ast}(-{\bm k})
\end{array}
\right),
\eeq
stand for the eigenfunctions of the upper/lower energy branches $E_{\pm}({\bm k})$. 

For an incident electron beam with an incident energy $E$, the tunneling conductance is
\beq
\sigma^{\rm S}(E) = \sum _s \left\langle \sigma^{\rm S}_{s}(E,\hat{\bm k})\right\rangle _{\hat{\bm k}},
\eeq
where $\sigma^{\rm S}_{s}(E,\hat{\bm k})$ is the angle-resolved tunneling conductance of incident electrons with spin $s$ and incident wavevector ${\bm k}$, given by
\beq
\sigma^{\rm S}_{s}(E,\hat{\bm k})
= 1 + \sum _{s^{\prime}}\left[ 
|a_{ss^{\prime}}(E,\hat{\bm k})|^2 - |b_{ss^{\prime}}(E,\hat{\bm k})|^2
\right].
\eeq

The coefficients, $a$, $b$, $c$, and $d$, are determined so as to follow the boundary conditions at the interface ($z=0$),
\begin{gather}
{\bm \psi}^{\rm N}(x,y,z\rightarrow 0_-) = {\bm \psi}^{\rm S}(x,y,z\rightarrow 0_+),  \\
\frac{\partial{\bm \psi}^{\rm N}({\bm r})}{\partial z}\bigg|_{z\rightarrow 0} - 
\frac{\partial{\bm \psi}^{\rm S}({\bm r})}{\partial z}\bigg|_{z\rightarrow 0} = \frac{2mH\psi(x,y,0)}{\hbar^2}.
\end{gather}
Analytic expressions for the conductance coefficients are obtained by solving the continuity conditions as~\cite{bohloul}
\begin{gather}
a_{s^{\prime}s} = k^2_z \left[ M^{-1}\right]_{ss^{\prime}}, \\
b_{s^{\prime}s} = -ik_z \left[
(Z\hat{v}\hat{u}^{\ast-1}+ Y\hat{u}\hat{v}^{\ast-1})M^{-1}
\right]_{ss^{\prime}}
-\delta _{ss^{\prime}}
\end{gather}
where we have introduced $Z\equiv mH/\hbar^2 k_{\rm F}$, $Y=Z+ik_z$,
$M = Z^2 \hat{v}\hat{u}^{\ast-1}
+ (Z^2+k^2_z)\hat{u}\hat{v}^{\ast-1}$,
and used abbreviation $\hat{u}\equiv \hat{u}({\bm k})$, $\hat{v}\equiv \hat{v}(-\underline{\bm k})$,
$\hat{u}^{\ast}\equiv \hat{u}^{\ast}(\underline{\bm k})$, and $\hat{v}^{\ast}\equiv \hat{v}(-{\bm k})$.

The conductance coefficients, $a_{ss^{\prime}}$ and $b_{ss^{\prime}}$, are functions of the barrier potential $Z$, the bias voltage $E$ and the gap function. The bias voltage is the energy eigenvalue of the solution. In unitary superconductors it appears in the expressions $\varepsilon = \sqrt{E^2 - |{\bm d}({\bm k})|^2}$ for $E^2 >|{\bm d}({\bm k})|^2$ and $\varepsilon = i \sqrt{|{\bm d}({\bm k})|^2-E^2}$ for $E^2 < |{\bm d}({\bm k})|^2$. The transformation matrices, $\hat{u}(E,\hat{\bm k})$ and $\hat{v}(E,\hat{\bm k})$, are obtained by replacing $E({\bm k})$ by the incident energy (bias) $E$, and $\hat{u}^{\ast}$ and $\hat{v}^{\ast}$ are evaluated by complex-conjugating all numbers except $\varepsilon ({\bm k})$. For nonunitary case, the energy $E_+$ ($E_-$) in the first and third (second and fourth) terms in Eq.~\eqref{eq:psis} is replaced by the incident energy $E$. To this end, the matrices for nonunitary case are given by
\begin{widetext}
\begin{align}
\hat{u}(E,{\bm k}) =& Q\left\{
\sqrt{\frac{E+\sqrt{E^2-|{\bm d}|^2-|{\bm q}|^2}}{E}}
(|{\bm q}|+{\bm q}\cdot {\bm \sigma})(\sigma _0 + \sigma _z) 
+ \sqrt{\frac{E + \sqrt{E^2-|{\bm d}|^2+|{\bm q}|^2}}{E}}
(|{\bm q}|-{\bm q}\cdot {\bm \sigma})(\sigma _0 - \sigma _z)
\right\}, \\
\hat{v}(E,{\bm k}) =& -i\frac{Q}{\sqrt{E}}\left\{
\frac{[|{\bm q}|{\bm d}-i({\bm q}\times {\bm q})]\cdot{\bm \sigma}\sigma _y}
{\sqrt{E+\sqrt{E^2-|{\bm d}|^2-|{\bm q}|^2}}}
(\sigma _0 + \sigma _z) 
+\frac{[|{\bm q}|{\bm d}+i({\bm d}\times {\bm q})]\cdot{\bm \sigma}\sigma _y}
{\sqrt{E+\sqrt{E^2-|{\bm d}|^2+|{\bm q}|^2}}}
(\sigma _0 - \sigma _z)
\right\},
\end{align}
with $Q({\bm k})\equiv [8|{\bm q}|(|{\bm q}|+q_z)]^{-1/2}$ and ${\bm q}\equiv i {\bm d}\times {\bm d}^{\ast}$.
\end{widetext}

\subsection{Cyclic state}

We first display the surface density of state in Fig.~\ref{fig:sdos2}, which is obtained by averaging the angle-resolved surface density of states over the Fermi surface,
\beq
{N}_{\rm S}(E) = \langle {N}(\hat{\bm k},z=0;E)\rangle _{\hat{\bm k}}.
\label{eq:dosk}
\eeq
In Fig.~\ref{fig:sdos2}, we also present the density of states in bulk cyclic states. The bulk density of states possesses two characteristic energies, $\Delta _{-}$ and $\Delta _{+}$, denoted by broken arrows. The former (latter) corresponds the maximal energy gap in the $E_-$ ($E_+$) quasiparticle branch. The inner gap within $|E|\lesssim 0.3\Delta $ represents the point nodal structure of $E_-({\bm k})$, while the coherence peak at $|E|=\Delta $ is attributed to the full gap structure of $E_+({\bm k})$. 

The surface density of states for $\vartheta/\pi = 0.1$ has two peaks at $E\approx 0.2 \Delta$ which are the van Hove singularities associated with the bended dispersion of the gapless surface bound states along the nodal direction (see Fig.~\ref{fig:sdos1}(a)). In consistent with the change of the surface dispersion, the split peaks shift to $E=0$ with increasing $\vartheta$ and merge to the zero energy peak. For the [111] surface, the surface density of states has a single pronounced peak at $E=0$. 

\begin{figure}[t!]
\includegraphics[width=85mm]{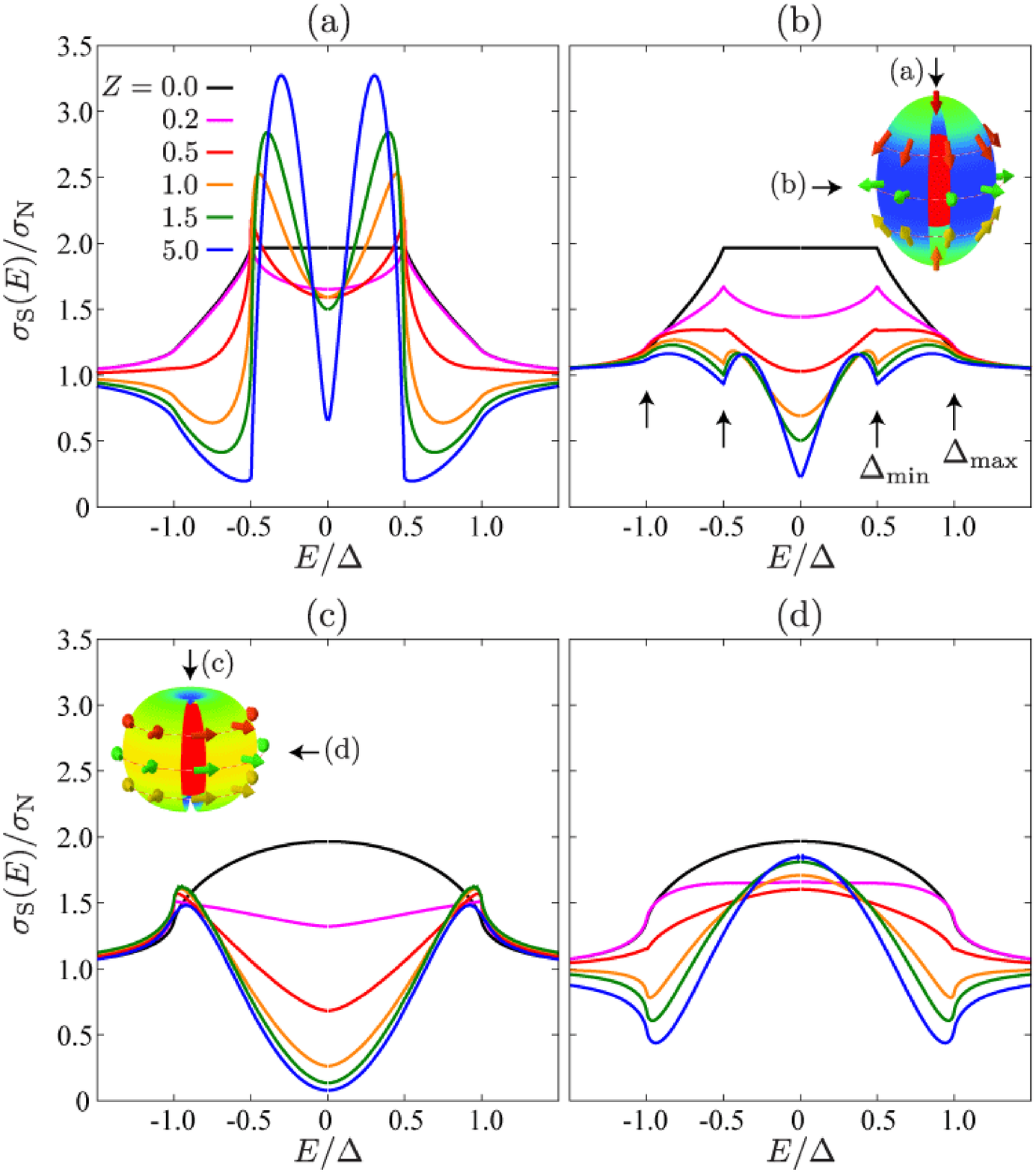}
\caption{Normalized tunneling conductance $\sigma^{\rm S}/\sigma^{\rm N}$ in uniaxial nematic state (a,b) and biaxial nematic state (c,d) for various the barrier potential $Z$: (a,c) the [001] surface and (b,d) the [110] surface.}
\label{fig:sdos3}
\end{figure}

The bottom panels in Fig.~\ref{fig:sdos2} show the normalized tunneling conductance $\sigma^{\rm S}(E)/\sigma^{\rm N}$ for cyclic $p$-wave states for various $\vartheta$. It is clearly seen that the results with a high potential barrier $Z$, corresponding to a low transparent interface, reveals the evolution of the surface density of states from the split peak originating in the van Hove singularities to the sharp zero energy peak associated with Majorana arcs. 
The pronounced zero bias conductance peak is attributed to the evolution of the Majorana arcs and is protected by the $P_2$ symmetry and Chern number. Hence, it is robust against the Rashba spin-orbit coupling on surface. In the case of low $Z$, corresponding to the high transparency, the peak structure around zero bias is smeared out and the evolution of surface states is not detectable. 

\subsection{Uniaxial/biaxial nematic states}

The degenerate scenario which was recently proposed in Ref.~\onlinecite{shimizu17} explains the multiple superconducting phases of U$_{1-x}$Th$_x$Be$_{13}$ in the basis of the $E_u$ irreducible representation of the $O_h$ symmetry. The cyclic state is consistent to the broken time reversal symmetry observed in the lower $T$ phase within $0.019\le x \le 0.045$, while the uniaxial (biaxial) nematic state occupies the phase in $ x \le 0.019$ (the higher $T$ phase within $0.019\le x \le 0.045$).

In Fig.~\ref{fig:sdos3}, therefore, we plot the normalized tunneling conductance in uniaxial and biaxial nematic states. The order parameters of the uniaxial and ($D_4$) biaxial nematic states are obtained from Eq.~\eqref{eq:dvec} with $(\eta _1, \eta _2) = (1,0)$ and $(0,1)$, respectively. For $(1,0)$, the quasiparticle gap is uniaxially elongated along the ${\bm c}$ direction and possesses two distinct gaps, $\Delta _{\rm max} = \Delta $ along ${\bm c}$ and $\Delta _{\min}=\Delta/2$ along $\hat{\bm a}$ and $\hat{\bm b}$. In contrast to the Weyl points in the cyclic state, two point nodes at ${\bm k}=\pm k_{\rm F}\hat{\bm c}$ in the biaxial nematic state are protected by the mirror reflection plane.~\cite{kobayashiPRB14}

The nematic states are categorized to the three-dimensional DIII topological class and the $4\!\times\! 4$ matrix, $\mathcal{H}({\bm k})$, is subject to Eq.~\eqref{eq:phs} with $\mathcal{C}^2\!=\!+1$ and Eq.~\eqref{eq:trs} with $\mathcal{T}^2\!=\!-1$. Hence, $\mathcal{H}({\bm k})$ is parameterized by the four-dimensional spinor $\hat{\bm m}\!=\!(\hat{m}_1,\hat{m}_2,\hat{m}_3,\hat{m}_4) \!\in\! S^3$, as $\mathcal{H}({\bm k})\!=\! |E({\bm k})|\sum ^4_{j=1}\hat{m}_j({\bm k})\gamma _j$, where $\gamma _j$ denotes the Dirac $\gamma$ matrices which obey $\{\gamma _i,\gamma _j\}=2\delta _{ij}$. This indicates that $\hat{\bm m}({\bm k})$ is a projector that maps ${\bm k} \!\in\! S^3$ onto the spinor space $\hat{\bm m}\!\in\!S^3$. The topological invariant relevant to the fundamental group, $\pi _3 (S^3) \!=\! \mathbb{Z}$, is the winding number~\cite{mizushimaJPCM15,mizushimaJPSJ16}, 
\begin{align}
w_{\rm 3d} =  \int \frac{d^3{\bm k}}{12\pi^3}
\epsilon_{\mu\nu\eta}\epsilon_{ijkl}
\hat{m}_i
\partial_{k_\mu}\hat{m}_j
\partial_{k_\nu}\hat{m}_k
\partial_{k_\eta}\hat{m}_l,
\label{eq:w3d}
\end{align}
which is calculated as $w_{\rm 3d}=-1$ for $r\!\neq\! -1$ ($\mu,\nu,\eta\!=\! x,y,z$ and $i,j,k,l\!=\!1,\cdots,4$).
For the $D_4$ biaxial nematic state at $r\!=\!-1$, the point nodes can be removed by adding a small perturbation that unchanges the symmetries. As a result, the winding number can be calculated as $w_{\rm 3d}=-1$ for the $D_4$ biaxial nematic state. As pointed out in Ref.~\onlinecite{satoPRB10}, however, an ambiguity in choosing the perturbation makes $w_{\rm 3d}$ gauge-dependent. Only the parity of $w_{\rm 3d}$, $\nu \equiv(-1)^{w_{\rm 3d}}\in \{ -1,+1\}$, reamins gauge-invariant. Hence, the nontrivial $\mathbb{Z}_2$ number $\nu = -1$ indicates that the $D_4$ biaxial nematic state is topological. 

Owing to the nontrivial $\mathbb{Z}$ and $\mathbb{Z}_2$ invariants, both the uniaxial and biaxial nematic states in cubic superconductors are accompanied by a single gapless Majorana cone and topologically protected Fermi arc, respectively. Solving the Andreev equation $\mathcal{H}(k_x,k_y,-i\partial _z)\varphi _{k_x,k_y}(z) = E (k_x,k_y)\varphi _{k_x,k_y}(z)$ with the boundary condition $\varphi (z) = 0$, one obtains the dispersion of the gapless surface state for the uniaxial/biaxial nematic statse as~\cite{mizushimaPRB17,mizushimaPRB12}
\beq
E_{\rm surf}(k_x,k_y) = \sqrt{v^2_xk^2_x + v^2_yk^2_y},
\label{eq:Esurf}
\eeq
where $(k_x,k_y)$ denotes the momentum parallel to the surface. For the uniaxial nematic state, the fully isotropic Majorana cone with the velocities $v_x =v_y= \Delta _{\rm min}/k_{\rm F}$ appears on the [001] surface, while the gapless states show the anisotropic dispersion with $v_x = \Delta _{\rm max}/k_{\rm F}$ and $v_y = \Delta _{\rm min}/k_{\rm F}$ in the case of the [100] surface. The Majorana nature and magnetic anisotropy of the gapless surface states were discussed in Ref.~\onlinecite{mizushimaPRB17}.

In Figs.~\ref{fig:sdos3}(a) and \ref{fig:sdos3}(b), we plot the tunneling conductance in the uniaxial nematic state. The conductance profiles are essentially different from those in the Balian-Werthamer (BW) state, i.e., the $A_{1u}$ state in $O_h$ crystals, having the isotropic Majorana cone,~\cite{asanoPRB03} and reveals the anisotropy of the dispersion of surface Majorana fermions. The isotropic BW state is accompanied by the isotropic cone with $v_x=v_y=\Delta$ and the surface density of states is linear on $|E|$ for $|E|\ll \Delta$. The tunneling conductance for large $Z$ shows the M-shaped broad double-hump structure within $|E|\le \Delta$,~\cite{asanoPRB03} as shown in Fig.~\ref{fig:accidental} ($\beta =0$). In contrast, when the long axis of the elongated gap in the uniaxial nematic state is normal to the surface, i.e., the [001] surface, the gapless surface states are confined to $|E|<\Delta _{\min}= \Delta _{\rm max}/2$. This gives rise to the squeezing of the M-shaped double-hump peak of $\sigma _{\rm S}(E)/\sigma _{\rm N}$ within $|E|<\Delta _{\rm min}$ as seen in Fig.~\ref{fig:sdos3}(a). For the [110] surface, however, the anisotropic dispersion implies that the surface states are distributed to the wide range of the energy within $|E|<\Delta _{\rm max}$. As shown in Fig.~\ref{fig:sdos3}(b), this broadens the $\sigma _{\rm S}(E)/\sigma _{\rm N}$ and the squeezed double-hump peak disappears.

For the biaxial nematic state, we set the nodal direction to be normal to the [001] surface. In Fig.~\ref{fig:sdos3}(c), the tunneling conductance spectra on the [001] surface shows the absence of the characteristic surface structure. The [110] surface is parallel to the nodal direction. The surface states are dispersionless along the nodal direction ($k_x$) and linear on $k_y$, i.e., $E_{\rm surf}(k_x,k_y)= \Delta k_y/k_{\rm F}$ for the up spin and $E_{\rm surf}(k_x,k_y)= -\Delta k_y/k_{\rm F}$ for the down spin. The resulting spectra in Fig.~\ref{fig:sdos3}(d) shows the dome-like peak without a pronounced zero bias peak, which is similar to the tunneling conductance spectra for chiral $p_x+ip_y$ superconductors.~\cite{kashiwayaPRL11}

\subsection{Accidental scenario}

\begin{figure}[t!]
\includegraphics[width=85mm]{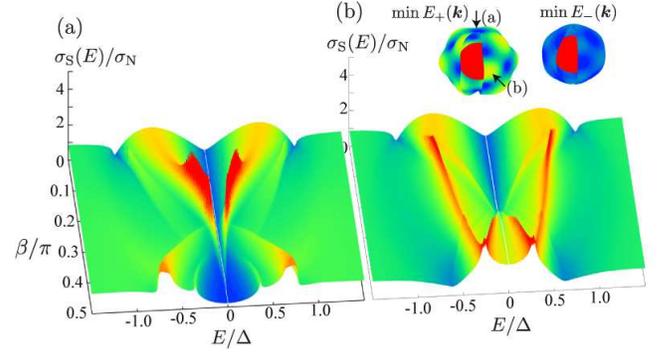}
\caption{Normalized tunneling conductance $\sigma^{\rm S}/\sigma^{\rm N}$ in the nonunitary $A_{1u}+iA_{2u}$ state: (a) the [001] surface and (b) the [110] surface. In all data, we fix $Z=5.0$. The inset shows the gap structures in the case of $\beta =\pi/5$.}
\label{fig:accidental}
\end{figure}

Another scenario for the superconducting gap of U$_x$Th$_{1-x}$Be$_{13}$ is the accidental scenario.~\cite{sigristPRB89,sigrist} 
This scenario assumes that two different one-dimensional irreducible representations of the $O_h$ group are accidentally nearly degenerate, and the ${\bm d}$ vector is obtained as a combination of two representations. Although the huge numbers of the combinations are possible, the recent experiment in Ref.~\onlinecite{shimizu17} can narrow down the possible gap symmetry. Following Ref.~\onlinecite{shimizu17}, we here consider the accidental degeneracy of the $p$-wave $A_{1u}$ and $f$-wave $A_{2u}$ states,
\beq
{\bm d}({\bm k}) = \Delta \left[ 
\cos\beta{\bm \Gamma}^{A_{1u}}+ i\sin\beta {\bm \Gamma}^{A_{2u}}\right],
\eeq
where $\beta \in [0,\pi/2]$.
The basis functions are given by ${\bm \Gamma}^{A_{1u}}= \hat{\bm a}\hat{k}_a + \hat{\bm b}\hat{k}_b+\hat{\bm c}\hat{k}_c$ and ${\bm \Gamma}^{A_{2u}}=\hat{\bm a}\hat{k}_a(\hat{k}^2_b-\hat{k}^2_c) + \hat{\bm b}\hat{k}_b(\hat{k}^2_c-\hat{k}^2_a) + \hat{\bm c}\hat{k}_c(\hat{k}^2_a-\hat{k}^2_b)$. The limit of $\beta = 0$ corresponds to the pure $A_{1u}$ state with the nodeless gap which is consistent to the full gap behavior in pure UBe$_{13}$ ($x=0$). For $\beta \in (0,\pi/2)$, the nonunitary chiral $A_{1u}\pm iA_{2u}$ state can explain both the broken time reversal symmetry and full gap behavior in $0<T<T_{\rm c2}$ at $x\sim 0.03$,~\cite{shimizu17} where $\beta$ remains as the fitting parameter. The pure $f$-wave state with $\beta = \pi/2$ occupies the higher $T$ phase in $0.019\le x \le 0.045$. Although the $A_{1u}+iA_{2u}$ state is nodeless as shown in the inset of Fig.~\ref{fig:accidental}, the $f$-wave $A_{2u}$ state has point nodes along the [100] and [111] directions.

Figure~\ref{fig:accidental} shows the tunneling conductance spectra in the nonunitary $A_{1u}+iA_{2u}$ state with various $\beta\in [0,\pi/2]$ for the [001] surface and the [110] surface (b). The $\beta=0$ case corresponds to the isotropic BW state, which shows the broad M-shaped double-hump structure irrespective of the surface orientation. For $\beta = \pi/2$, the spectrum on the [001] surface shows the $E^2$ dependence within $|E|\ll \Delta$, which reveals the point node along the [001] direction. The tunneling spectra for all $\beta \in [0,\pi/2]$ does not have any pronounced peak structure in the vicinity of the zero energy.

\section{Concluding remarks}

In this paper, we have discussed the symmetry and topology of surface states in superconductors with nonunitary cyclic pairing. The low energy physics is governed by itinerant Majorana fermions in the bulk, while gapless surface states show the evolution from a single cone to zero energy arcs under rotation of surface orientation. We have clarified that the gapless Majorana cone is protected solely by accidental spin-rotation symmetry, while the Majorana arcs are protected by two different topological invariants: the first Chern number originating in eight Weyl points at [111] direction and one-dimensional winding number associated with the combined symmetry of time reversal and mirror reflection. Hence, the gapless cone is fragile against the spin-orbit interaction.

Using the BTK theory, we have calculated tunneling spectra in the nonunitary cyclic state, the uniaxial/biaxial nematic states, and the $A_{1u}+iA_{2u}$ state for various surface orientations. By changing the surface orientation from the [001] direction to the [110] direction, in the nonunitary cyclic state, the tunneling conductance with a high barrier potential shows the evolution from the sharp double peak structure to a pronounced zero-bias conductance peak. The former reflects the van Hove singularities in the dispersion of Majorana cone and the latter is attributed to the existence of the zero energy surface Majorana arcs. Such a pronounced zero-bias conductance peak cannot be observed in the a the $A_{1u}+iA_{2u}$ representations, irrespective of the surface orientation. The cyclic ($A_{1u}+iA_{2u}$) state is the candidate for the broken time reversal symmetry state of U$_{1-x}$Th$_x$Be$_{13}$ ($0.019\le x \le 0.045$) in the degenerate $E_u$ (accidental) scenario.~\cite{shimizu17,sigristPRB89,sigrist} Hence, the tunneling spectroscopy can clearly capture the topologically protected surface states in nonunitary cyclic superconductors.

Lastly, we would like to mention that the pronounced zero bias conductance peak was observed in the UBe$_{13}$ superconductor-normal metal (Au) junction.~\cite{waltiPRL00}Niether the degenerate senario nor the accidental senario explains the characteristic spectra in the $x=0$ case. The BW ($A_{1u}$) state in the accidental senario shows the M-shaped double-hump conductance peak regardless of the surface orientation, while the enhancement or suppression of the M-shaped peak is realized in the uniaxial nematic state in the degenerate senario. The discrepancy might be attributed to the polycrystal of the pure UBe$_{13}$, where the tip radii of the Au tip are much larger than the average graine size. Our main outcomes may be useful for further tunneling spectroscopy measurements in high-quality single crystals. The discrepancy may also originate in the characteristic electrons of the normal states. For instance, it has been shown that the intertwining of surface Majorana fermions with surface states proper to topological insulators gives rise to the transition of the dispersion of the surface state and a pronounced zero-bias conductance peak may appear even in a fully gapped topological state.~\cite{hsiehPRL12,yamakagePRB12,haoPRB11,mizushimaPRB14v2} Hence, the discrepancy may be resolved by taking into account the more realistic information of the material such as the topology of the Fermi surface~\cite{takegahara,maehira} and so on.



\begin{acknowledgments} 

We thank K. Machida for bringing new pairing scenario of heavy-fermion superconductor U$_{1-x}$Th$_x$Be$_{13}$ to our attention.
This work was supported by Japan Society for the Promotion of Science (JSPS) (Grants No.~JP16K05448 (T.M.) and No.~JP16H03948 (M.N.)) and ``Topological Materials Science'' (Grant No.~JP15H05855) and ``Nuclear Matter in Neutron Stars Investigated by Experiments
and Astronomical Observations'' (Grant No.~JP15H00841) KAKENHI on innovation areas from Ministry of Education, Culture, Sports, Science and Technology (MEXT). The work of M.N. is also supported in part by the MEXT-Supported Program for the Strategic Research Foundation at Private Universities ``Topological Science'' (Grant No.~S1511006).

\end{acknowledgments}

\appendix

\section{Quasiclassical theory}

The quasiclassical propagator $g\equiv{g}(\hat{\bm k},{\bm r};\varepsilon _n)$ is governed by the transport-like equation. Following the procedure in Ref.~\onlinecite{serene}, one obtains the quasiclassical transport equation from the Gor'kov equation as
\beq
\left[i\varepsilon _n {\tau}_z - {v}(\hat{\bm k},{\bm r})
- \underline{\Delta}(\hat{\bm k},{\bm r}), 
{g}
\right] + i {\bm v}_{\rm F} \cdot{\bm \nabla}
{g}=0.
\label{eq:eilen}
\eeq
The Fermi velocity is defined as ${\bm v}_{\rm F}(\hat{\bm k}) \!=\! \partial \varepsilon _0({\bm k})/\partial {\bm k}|_{{\bm k}=k_{\rm F}\hat{\bm k}}$. The external potential, $\underline{v}(\hat{\bm k},{\bm r})$, is given with a magnetic Zeeman field as
$\underline{v}(\hat{\bm k},{\bm r}) = - \frac{1}{1+F^{\rm a}_0}\frac{1}{2}\gamma H_{\mu}
\sigma _{\mu} \oplus \sigma^{\rm T}_{\mu}$, 
where $F^{\rm a}_0$ is the Fermi liquid parameter. We here omit the self-energies associated with the Fermi liquid corrections. The off-diagonal component of the quasiclassical self-energies is given as
\beq
\underline{\Delta}(\hat{\bm k},{\bm r}) 
= \left(
\begin{array}{cc}
0 & i{\bm \sigma}\cdot{\bm d}(\hat{\bm k},{\bm r})\sigma _y \\
i\sigma _y{\bm \sigma}\cdot{\bm d}^{\ast}(\hat{\bm k},{\bm r})& 0
\end{array}
\right). 
\eeq
The quasiclassical transport equation (\ref{eq:eilen}) is a first-order ordinary differential equation along a trajectory in the direction of ${\bm v}_{\rm F}(\hat{\bm k})$. To obtain a unique solution for $g$, Eq.~(\ref{eq:eilen}) must be supplemented by the normalization condition,
\beq
[ \underline{g} (\hat{\bm k},{\bm r};\varepsilon _n)]^2 = -\pi^2.
\label{eq:norm}
\eeq

The order parameters for the $E_u$ representation is determined by solving the gap equation
${\bm d} (\hat{\bm k},{\bm r}) = \sum _{m=1,2} \eta _m ({\bm r}) 
{\bm \Gamma}^{E_u}_m(\hat{\bm k})$. The self-consistent ${\bm d}$-vector field is obtained from the anomalous propagator by solving the gap equation, 
$
d_{\mu}(\hat{\bm k},{\bm r}) = T\sum _{n}\langle 
V_{\mu \nu}(\hat{\bm k},{\bm k}^{\prime})f_{\mu}(\hat{\bm k}^{\prime},{\bm r};\varepsilon _n)\rangle _{\hat{\bm k}^{\prime}}
$.
We use the following abbreviation for the average over the Fermi surface, 
$\langle\cdots\rangle _{\hat{\bm k}} = \frac{1}{\mathcal{N}_{\rm F}}
\int \frac{d\hat{\bm k}}{(2\pi)^3 |{\bm v}_{\rm F}(\hat{\bm k})|}\cdots$, and $\sum _n$ denotes the Matsubara sum with the cutoff energy $E _{\rm c}$.
Assuming the separable form of the pairing interaction, $V_{\mu \nu}(\hat{\bm k},\hat{\bm k}^{\prime})=-\sum _m g_m \Gamma _{m,\mu}(\hat{\bm k})\Gamma^{\ast}_{m,\nu}(\hat{\bm k}^{\prime})$, one obtains the self-consistent equation for $\eta _{m}({\bm r})$ as 
\beq
\eta _m ({\bm r}) = -g_m T\sum _n \left\langle {\bm \Gamma}^{\ast}_{m}(\hat{\bm k})\cdot{\bm f}(\hat{\bm k},{\bm r};\varepsilon _n)\right\rangle.
\eeq
%
%
%
The coupling constant ($g_m>0$) is determined by the transition temperature $T^{(m)}_{\rm c}$ through the linearized gap equation at the superconducting critical temperature $T=T^{(m)}_{\rm c}$, $g^{-1}_m = \frac{1}{3}\sum _{|\varepsilon _n|<\varepsilon _{\rm c}} \frac{1}{|(2n+1)|}$. For simplicity, we set $T^{(m=1)}_{\rm c}=T^{(m=2)}_{\rm c}=T_{\rm c}$.

The numerical integration of the quasiclassical equation with the normalization condition can be simplified by introducing a parametrization for the propagator~\cite{eschrigPRB00,eschrigPRB99,nagatoJLTP93}
\beq
g = -i\pi{N} \left(
\begin{array}{cc}
1+\gamma\bar{\gamma} &  2 \gamma \\
-2\bar{\gamma} & -1 - \bar{\gamma}\gamma
\end{array}
\right),
\eeq
where ${N}\equiv (1-\gamma\bar{\gamma})^{-1}\oplus(1-\bar{\gamma}\gamma)^{-1}$. This parameterization satisfies the normalization condition by construction and reduces the number of independent components. By using the parameterization, Eq.~\eqref{eq:eilen} is generally mapped onto the Riccati-type differential equation
\begin{gather}
i{\bm v}_{\rm F}\cdot{\bm \nabla}\gamma - \gamma \bar{\Delta} \gamma + (i\varepsilon _n- \nu )\gamma - \gamma (-i\varepsilon _n - \bar{\nu})
+ \Delta 
=0, \\
i{\bm v}_{\rm F}\cdot{\bm \nabla}\bar{\gamma} - \bar{\gamma} {\Delta} \bar{\gamma} + (-i\varepsilon _n- \bar{\nu} )\bar{\gamma} - \bar{\gamma} (i\varepsilon _n - {\nu})
+ \bar{\Delta} 
=0,
\label{eq:riccati}
\end{gather}
with $\Delta \equiv i{\bm \sigma}\cdot{\bm d}\sigma _y$ and $\bar{\Delta} \equiv i\sigma _y {\bm \sigma}\cdot{\bm d}$. The Riccati amplitudes obey the relation, $\bar{\gamma}(\hat{\bm k},{\bm r};\varepsilon _n) = \gamma^{\ast}(-\hat{\bm k},{\bm r};\varepsilon _n)$.

For quasiparticle momentum $\hat{\bm k}$, the Riccati equations for $\gamma (\hat{\bm k})$ and $\bar{\gamma} (\hat{\bm k})$ are numerically stable along the quasiclassical forward ($\hat{\bm k}$) and backward ($-\hat{\bm k}$) trajectories with an initial value, respectively. We perform the numerical integration of Eq.~\eqref{eq:riccati} with the fourth-order Runge-Kutta method from the homogenous solution at $z=\infty$. For nonunitary state with ${\bm q}\equiv i{\bm d}\times {\bm d}^{\ast}\neq {\bm 0}$, the homogeneous solution with constant ${\bm d}$, $\gamma _{\mu}\equiv \frac{1}{2}{\rm tr}(-i\sigma _y \sigma _{\mu}\gamma)$, is given by
\begin{align}
\gamma _{\mu}(\hat{\bm k},z=\infty;\tilde{\varepsilon}) 
=& -\frac{|{\bm d}(\hat{\bm k})|^4}{|{\bm d}(\hat{\bm k})\cdot{\bm d}(\hat{\bm k})|^2}\nn \\
&\times \frac{d_{\mu}(\hat{\bm k})+i[{\bm d}(\hat{\bm k})\times {\bm q}(\hat{\bm k})]_{\mu}/|{\bm d}(\hat{\bm k})|}{\tilde{\varepsilon}+is\sqrt{|{\bm d}(\hat{\bm k})|^2-\tilde{\varepsilon}^2}}.
\end{align}
where $s=+1$ for ${\rm Im}\tilde{\varepsilon}>0$ and $s=-1$ for ${\rm Im}\tilde{\varepsilon}<0$. We here set $\tilde{\varepsilon} = i\varepsilon _n$ for the Matsubara propagator $\gamma (\varepsilon _n)$ and $\tilde{\varepsilon} = E \pm i0_+$ for the retarded and advanced propagators $\gamma^{\rm R,A}(E) = \gamma (\varepsilon _n \rightarrow -iE + 0_+)$. We impose the boundary condition on the $4\times 4$ quasiclassical propagator $g(\hat{\bm k},{\bm r};\varepsilon _n)$ as
\beq
\gamma(\hat{\bm k},{\bm r}_{\rm surf};\varepsilon _n) = \gamma (\underline{\hat{\bm k}},{\bm r}_{\rm surf};\varepsilon _n),
\label{eq:bc}
\eeq
and $\bar{\gamma}$ as well, which represent the specular scattering of quasiparticles on the surface.

\bibliography{cyclic.bib}

\end{document}